\documentclass[useAMS,usenatbib]{mnras}
\usepackage{amsmath}
\usepackage{float}
\usepackage{graphicx}
\usepackage{txfonts}
\usepackage{indentfirst}
\usepackage{color}
\usepackage{ulem}
\usepackage{hyperref}
\usepackage[Symbolsmallscale]{upgreek}

\newcommand{\eq}[1]{\begin{equation}  #1 \end{equation}}
\newcommand{\eqs}[1]{\begin{equation} \begin{split} #1 \end{split} \end{equation}}

\newcommand{\br}[1]{\left( #1 \right)}

\newcommand{\bb}[1]{\left[ #1 \right]}

\newcommand{\dd}{{\rm d}}

\def\apj{ApJ}
\def\apjl{ApJ}
\def\apjs{ApJS}
\def\aap{A\&A}

\def\mnras{MNRAS}
\def\nat{Nature}

\def\physrep{Phys.~Rep.}

\def\aapr{A\&A~Rev.}

\title[Dynamical heating of the ICM]{Dynamical heating of the X-ray emitting intracluster medium: the roles of merger shocks and turbulence dissipation}
\author[Shi et al.]
{Xun Shi$^{1}$\thanks{E-mail: xun@ynu.edu.cn},
Daisuke Nagai$^{2}$,
Han Aung$^{2}$,
Andrew Wetzel$^{3}$
\\
$^{1}$South-Western Institute for Astronomy Research (SWIFAR), Yunnan University, 650500 Kunming, P. R. China\\
$^{2}$Department of Physics, Yale University,
New Haven, CT 06520, USA \\
$^{3}$Department of Physics, University of California, Davis, CA USA
}

\begin{document}

\maketitle

\begin{abstract}
The diffuse plasma inside clusters of galaxies has X-ray emitting temperatures of a few keV. The physical mechanisms that heat this intracluster medium (ICM) to such temperatures include the accretion shock at the periphery of a galaxy cluster, the shocks driven by merger events, as well as a somewhat overlooked mechanism -- the dissipation of intracluster turbulent motions. We study the relative role of these heating mechanisms using galaxy clusters in Lagrangian tracer particle re-simulations of the Omega500 cosmological simulation. We adopt a novel analysis method of decomposing the temperature increase at each time step into the contribution from dissipative heating and that from adiabatic heating. In the high-resolution spatial-temporal map of these heating rates, merger tracks are clearly visible, demonstrating the dominant role of merger events in heating the ICM. The dissipative heating contributed by each merger event is extended in time and also occurs in the rarefaction regions, suggesting the importance of heating by the dissipation of merger-induced turbulence. Quantitative analysis shows that turbulence heating, rather than direct heating at merger shocks, dominates the temperature increase of the ICM especially at inner radii $r < r_{\rm 500c}$. In addition, we find that many merger shocks can propagate with almost constant velocity to very large radii $r \gg r_{\rm 500c}$, some even reach and join with the accretion shock and becoming the outer boundary of the ICM. Altogether, these results suggest that the ICM is heated more in an `inside-out' fashion rather than `outside-in' as depicted in the classical smooth accretion picture.
\\
\end{abstract}


\section{Introduction}
Galaxy clusters contain intracluster medium (ICM), which consists of baryonic gas with X-ray emitting thermal temperatures. There is a consensus that this high temperature originates predominantly from gravitational energy released during the assembly of galaxy clusters' large gravitational potential well as a result of cosmic structure growth \citep[][for a recent review]{walker19}. Non-gravitational processes such as radiative cooling and feedback from SNe and AGN modify the temperature only in the central regions of the ICM \citep[e.g.,][]{barnes18,tremmel19} and have limited influence on ICM thermodynamics when we consider the whole cluster region.

The detailed mechanism of the conversion from gravitational binding energy to the thermal energy of the ICM, e.g., when, where and how is the ICM heated, is less clear. In the classical smooth accretion picture of ICM assembly \citep[e.g.,][]{tozzi01, voit03}, the accretion shock at the periphery of the ICM plays a dominant role in heating the ICM. In this picture, if non-gravitational heating is neglected, ICM gas entropy is gained only at the accretion shock when the gas falls smoothly from the intergalactic space onto a galaxy cluster. Subsequently, the gas gradually moves inward due to its residue infall velocity and its temperature further increases due to adiabatic compression, during which process the entropy is kept fixed.

On the other hand, it is well-known that merger events, especially major mergers, have a great impact on the heating of the ICM.
Mergers affect ICM heating in several ways. Firstly, mergers correspond to a different way of galaxy cluster mass growth other than smooth accretion.
Gas carried into the ICM by merging objects does not go through the accretion shock. Instead, it maintains a high velocity in the ICM, generating merger shocks while getting stripped and mixed with the ICM \citep[e.g.,][]{ricker01, paul11, ha18, zhangcy19b}. These merger shocks reveal themselves in X-ray observations as sharp surface brightness edges and temperature discontinuities, and sometimes also in radio observations as radio relics \citep[e.g.,][]{markevitch07, feretti12, bykov15, vanweeren19}. As they propagate through the galaxy cluster, the merger shocks heat the ICM \citep{sar02, mcc07, zuhone11}. Although they have typically low Mach numbers of $2-4$, much lower than that of the accretion shock, the heating from these merger shocks in terms of temperature increase may still
be significant.

Secondly, merger events generate turbulent gas motions in the ICM, as demonstrated by hydrodynamic numerical simulations \citep[e.g.,][]{norman99, iapichino08, nelson12, nagai13, miniati14, iapichino17, vazza12, vazza17, valdarnini11, valdarnini19}.
These turbulent gas motions provide non-thermal pressure \citep{lau09,bat12,nelson14b,shi14,shi15} that biases the cluster mass estimate based on hydrostatic equilibrium \citep{rasia06,nagai07,piffaretti08,bat12,lau13,shi16,biffi16,henson17,ansarifard20}, mix the chemicals in the ICM \citep{rus10}, and generate intracluster magnetic fields \citep{cho14, beresnyak16, vazza18, donnert18}. They can carry a significant amount of kinetic energy, which eventually dissipates to heat. Compared to the merger shock heating which happens instantaneously at shock fronts, this merger-induced turbulence heating corresponds to delayed heating that is extended in time. Also, the characteristic time scale of turbulence dissipation heating in the ICM is radial dependent \citep[e.g.,][]{shi14, shi18}, which may play a significant role in shaping the temperature structure of the ICM \citep{avestruz16}.

The smooth accretion and the effect of the mergers represent two different aspects of ICM assembly. What are their roles in a unified picture of ICM heating? What is the relative contribution of merger shocks and turbulence dissipation? How do these depend on radius and time? Answering these questions is essential for understanding the structures of ICM properties and their connection to the large scale cosmic environment, which are among the key science questions of the upcoming X-ray and Sunyaev-Zel'dovich (SZ) programs \citep[see e.g.,][for recent reviews and references therein]{walker19,mroczkowski19}, such as eROSITA, XRISM, Athena, Lynx, Cosmic Web Explorer in X-ray and Simons Observatory, CMB-S4, CMB-HD in microwave. In this paper, we shall investigate these questions using hydrodynamical cosmological simulations of galaxy cluster formation.

Finding the answers is complicated by the fact that the ICM is a compressive media, and thus density variations also lead to adiabatic temperature variations apart from the irreversible heating from dissipative processes.
Luckily, these two sources of heating can be clearly separated by examining the entropy $K = P / \rho_{\rm gas}^{\gamma}$, with $P$ and $\rho_{\rm gas}$ being the pressure and density of the gas, and $\gamma$ being the adiabatic index. While dissipative heating increases entropy, adiabatic temperature variations keep entropy unaltered.

In this study, we use a tracer particle re-simulation of the Omega500 simulation, a large Eulerian cosmological simulation specially designed to study galaxy clusters \citep{nelson14}, to study the thermal history of the ICM. We focus on the sources, radial distribution and temporal distribution of the ICM heating.

\section{Decomposition of dissipative and adiabatic heating}
\label{sec:decomposition}
Gas entropy is a useful variable in revealing the ICM thermodynamical
history. We use it to
distinguish dissipative heating including heating by the accretion shock, merger shocks,
and the dissipation of turbulence, and adiabatic
heating/cooling caused by compression/rarefaction of the gas. Namely, we decompose the contribution to the ICM temperature increment into
\eq{
\label{eq:decomposition}
\dd T = \dd T_{\rm diss} + \dd T_{\rm ad}
}
for each tracer particle at each time step,
with dissipative heating $\dd T_{\rm diss}$ being the \textit{irreversible} heating associated with entropy increase,
\eq{
\label{eq:dTdiss}
\dd T_{\rm diss} = T \dd \ln K \,,
}
and adiabatic heating $\dd T_{\rm ad}$ denoting the heating/cooling from
density increase/decrease which keeps the gas entropy constant,
\eq{
\label{eq:dTad}
\dd T_{\rm ad} = (\gamma - 1) T \dd \ln \rho_{\rm gas} \,.
}

This decomposition (Eq.\;\ref{eq:decomposition}) is the key method of this paper. We shall use it in several different ways to explore different physics:
\begin{enumerate}
 \item  We access the importance of the adiabatic heating in the smooth accretion picture by comparing the average dissipative heating $\Delta T_{\rm diss}$ and adiabatic heating $\Delta T_{\rm ad}$ across the growth history of a galaxy cluster.

 \item To see where and when the ICM is heated, we visualize the spatial-temporal distribution of $\dd T_{\rm diss}$ and $\dd T_{\rm ad}$.

 \item The relative role of shock heating and heating by turbulence dissipation is evaluated by comparing the accumulated dissipative heating $\Delta T_{\rm diss}$ that occurred in the compression region of the ICM ($\dd \rho_{\rm gas} > 0$ i.e., $\dd T_{\rm ad} > 0$) and that in the rarefaction region of the ICM ($\dd \rho_{\rm gas} < 0$ i.e., $\dd T_{\rm ad} < 0$).

 \item We estimate the Mach number of the merger shocks by comparing $\dd T_{\rm diss}$ and $\dd T_{\rm ad}$ at merger shock fronts.
\end{enumerate}

The results of these analyses will be presented in Sections \ref{sec:thermal_history} to \ref{sec:merger_shock} respectively.

\section{Tracer Particle Simulation and the selected galaxy cluster}
\begin{figure}
\centering
    \includegraphics[width=0.38\textwidth]{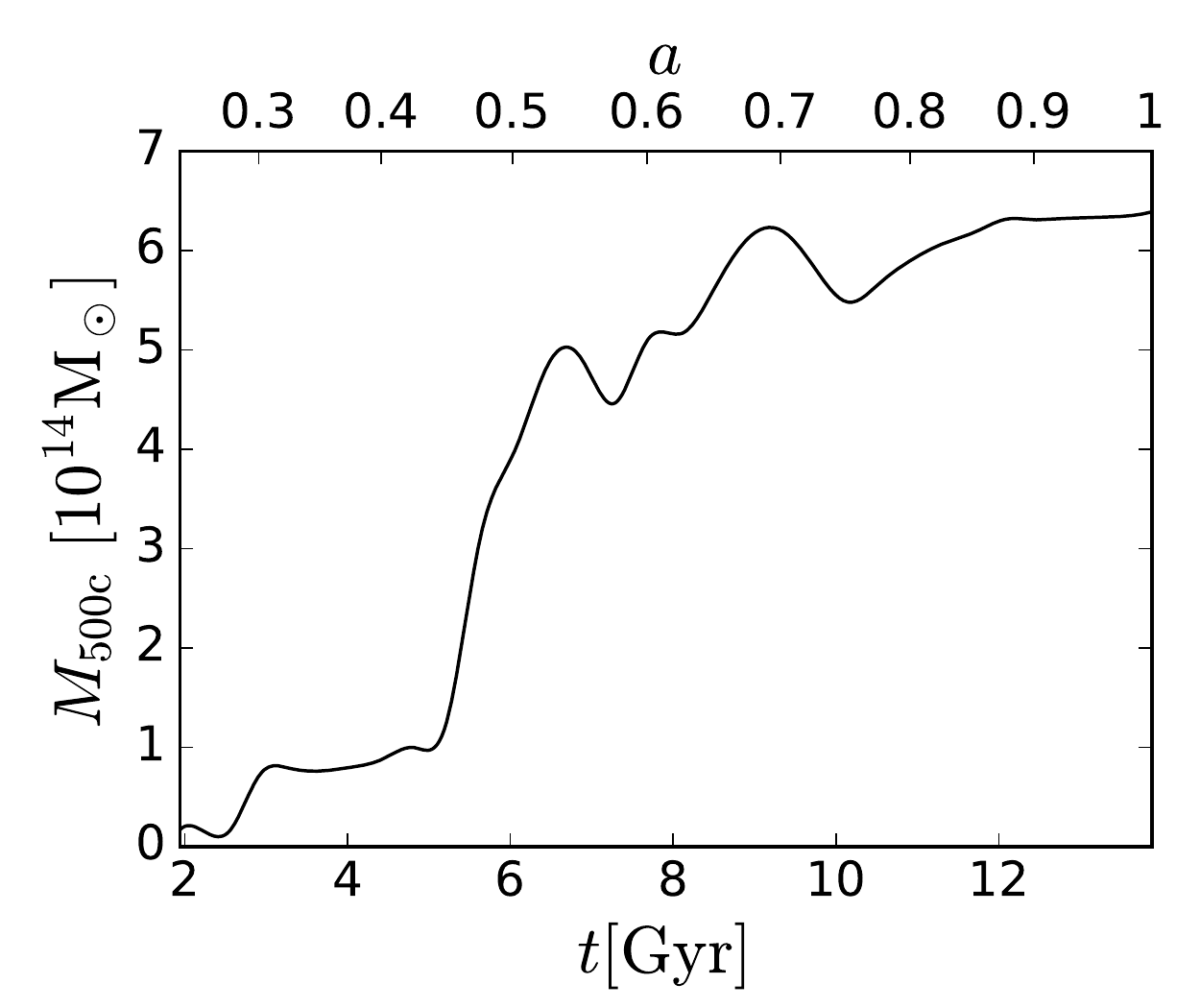}\\
    \caption{Mass growth history of the selected galaxy cluster. $M_{\rm 500c}$ is the mass enclosed in a radius in which the mean matter density is 500 times that of the critical density of the universe, $t$ is cosmic time, and $a$ is the cosmic scale factor.}
\label{fig:MAH}
\end{figure}

\begin{figure*}
\centering
    \includegraphics[width=0.98\textwidth]{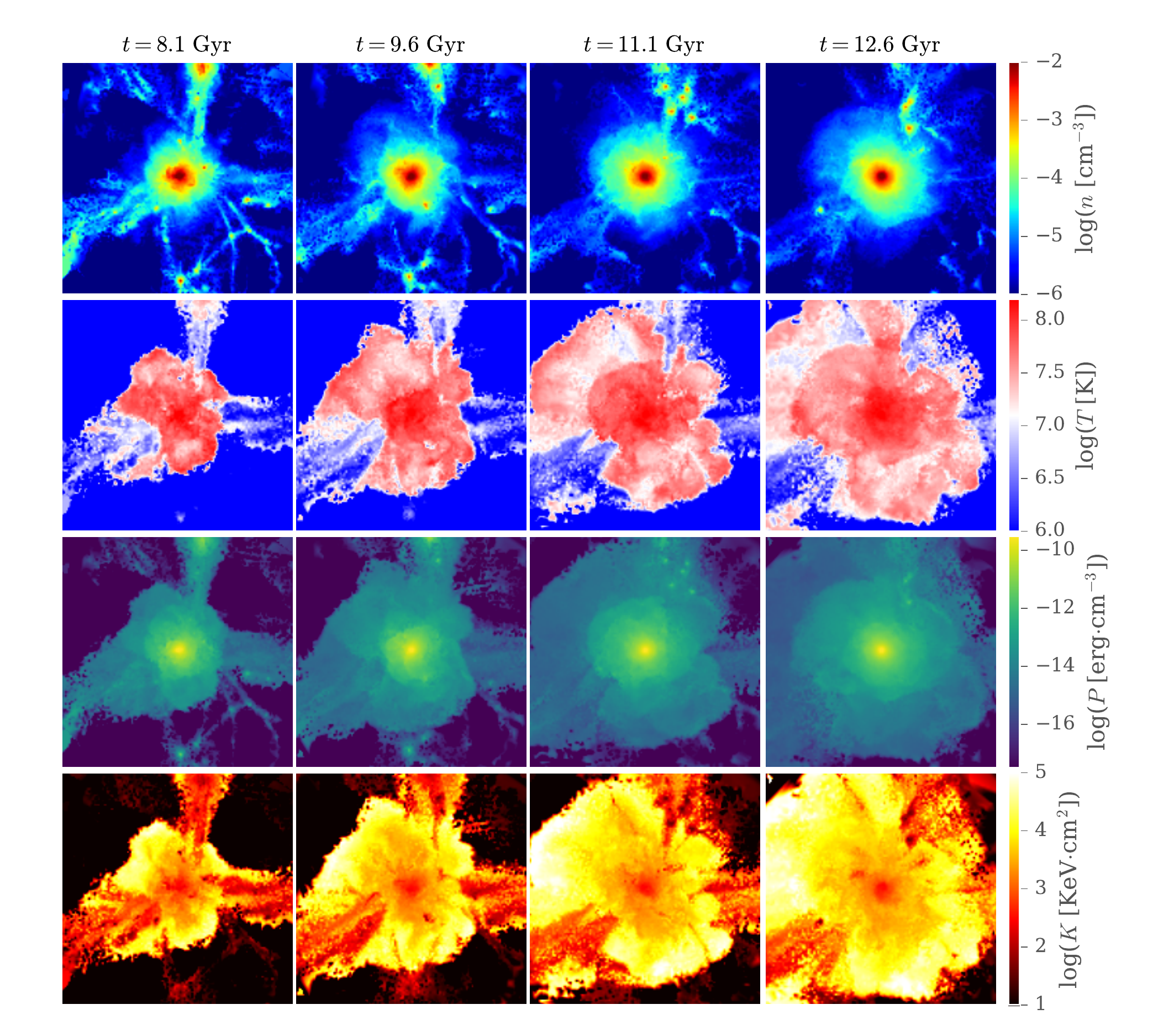}\\
    \caption{Gas number density, temperature, pressure, and entropy distributions in a $16 \times 16$ Mpc (physical) slice centered on the selected cluster at four epochs. The thickness of the slice is limited by an opening angle of $\uppi/10$, (i.e., we take $9\uppi/20 < \theta <11\uppi/20$) with respect to the z-axis being perpendicular to the presented slice.}
\label{fig:image}
\end{figure*}

\begin{figure}
\centering
    \includegraphics[width=0.37\textwidth]{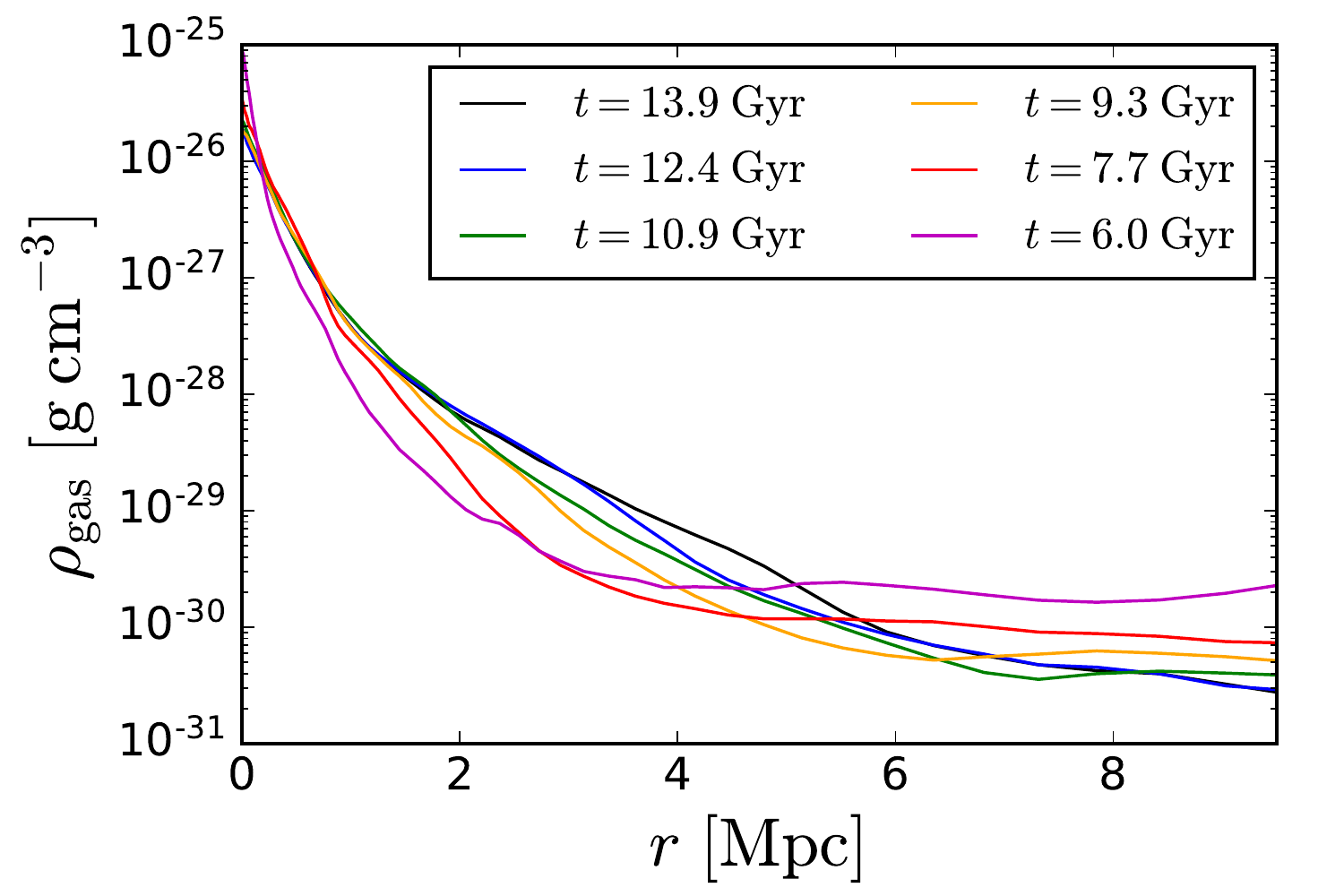}\\
    \includegraphics[width=0.37\textwidth]{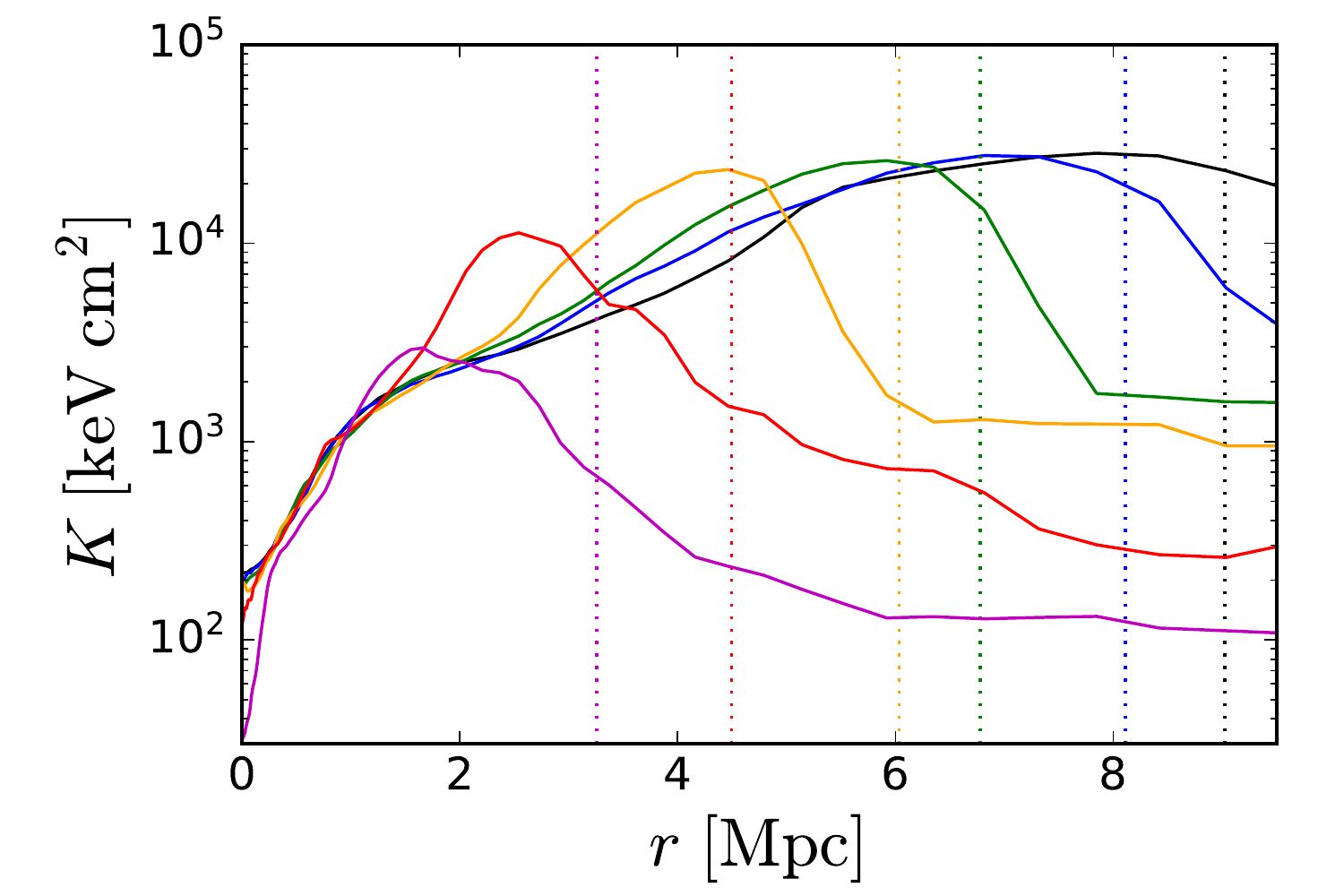} \\
    \includegraphics[width=0.37\textwidth]{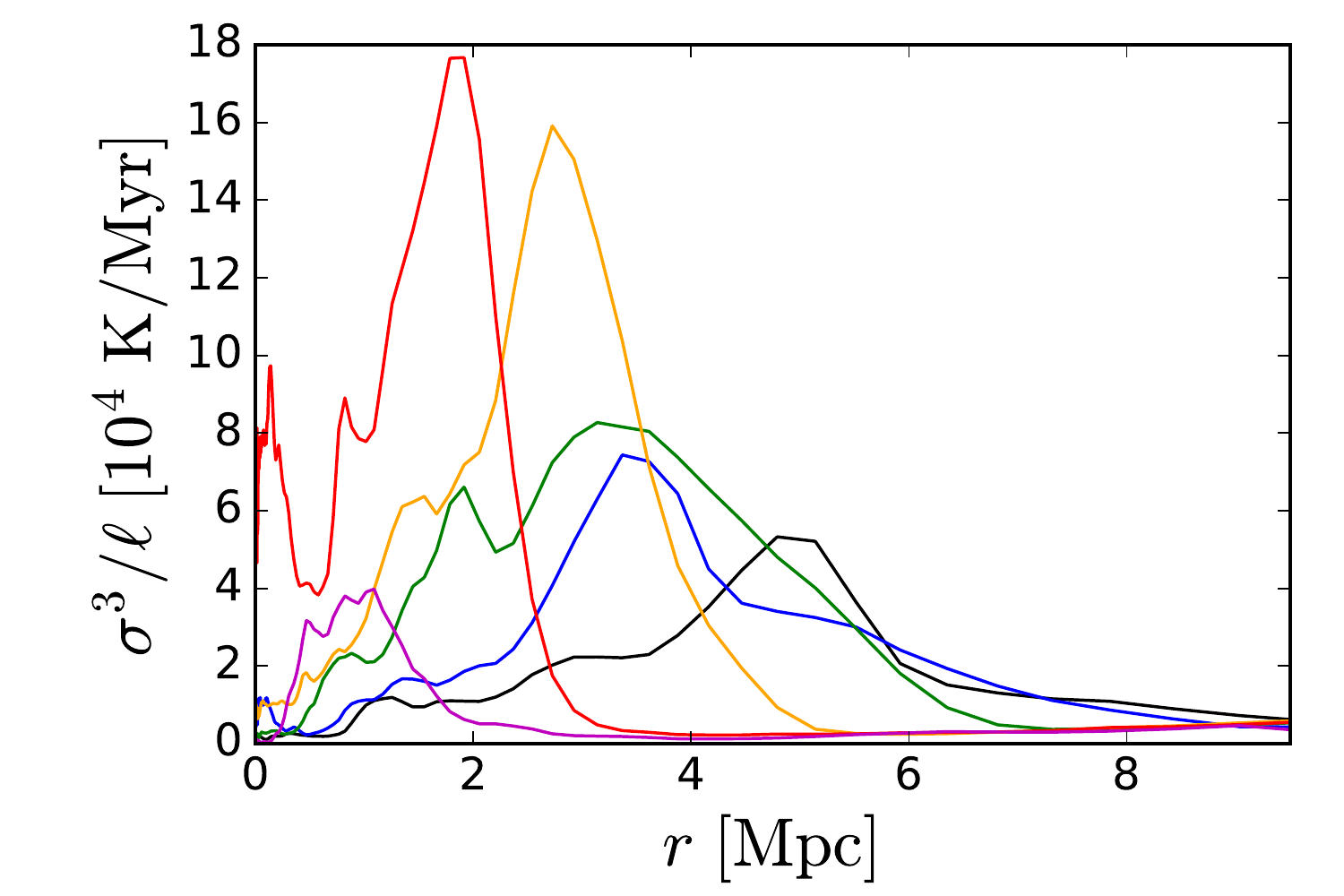}
    \caption{\textsl{Top panel:} Time evolution of gas density profile around the selected galaxy cluster. Here the profiles are computed by averaging over shells of the original Eulerian simulation, and the contribution from sub-structures (clumps) has been removed using the method introduced in \citet{zhur13b}. \textsl{Middle panel:} Evolution of entropy profile.
    The vertical dashed lines mark the volume average of accretion shock positions at the corresponding cosmic time, with the accretion shock position in a single direction computed as the outer most entropy jump along that sightline. \textsl{Bottom panel:} A rough estimate of turbulence heating rate made with the ICM radial velocity dispersion $\sigma$ measured from the Eulerian simulation and a fixed $\ell = 200$ kpc as a typical scale where the turbulence energy peaks in the ICM (cf. Fig.\;4 in \citealt{shi18}).}
\label{fig:profile_evo}
\end{figure}

We base our study on the non-radiative run of the  Omega500 simulation \citep{nelson14}, a large cosmological Eulerian simulation specifically designed to study the evolution of galaxy clusters. A second simulation with finer resolution is performed in regions surrounding galaxy clusters selected from the initial simulation. This yields a large sample of highly resolved galaxy clusters at redshift zero (see \citealp{nelson14,nelson14b} for more information). The simulation is performed in a flat $\Lambda$CDM cosmology with
WMAP five-year (WMAP5) cosmological parameters: $\Omega_{\rm m} = 0.27$, $\Omega_{\rm b} = 0.0469$,  $h = 0.7$ and $\sigma_{\rm 8}= 0.8$, with $h$ being the reduced Hubble constant and $\sigma_{\rm 8}= 0.8$ the
mass variance within spheres of radius of $8 \;h^{-1} \rm Mpc$. The simulation box size is $500 \; h^{-1} \rm Mpc$ and the maximum comoving spatial resolution is  $3.8 \; h^{-1} \rm kpc$. Since our focus is on ICM heating over large radial scales, we do not discuss the $\sim 100$ kpc core region where radiative cooling could be important. Also, we do not include heating from feedback processes which can affect the inner part of the ICM, but instead concentrate on the heating from structure growth which dominantes the overall heating of the ICM. Using non-radiative simulations not only suffices for our purpose but also greatly simplifies the analysis and interpretation.

A Lagrangian perspective has a significant advantage over an Eulerian one in studying the thermal history of gas, because it tracks particle trajectories and thus can  separate heating and cooling from transport effects.
To enable a Lagrangian study of the ICM, we perform a re-simulation of the Eulerian simulation with a large number 
of Lagrangian tracer particles. The tracer particles are injected into the simulations at the initial condition with a number density proportional to the local gas density in regions surrounding galaxy clusters. Then the tracer particles are passively advected with time using the 3D velocity field of the Eulerian simulation, and their thermodynamical quantities of the single highest resolution gas cell that the particle belongs to are reported when outputting the tracer particle information.
The output of the tracer particle re-simulation is saved at a very fine time resolution of 20-30 Myr, which amounts to more than 400 snapshots between redshifts $z=4$ and $z=0$. More discussion of the tracer particle method can be found in Appendix\;\ref{app:tracer}.

We apply the decomposition described in Section\;\ref{sec:decomposition} to galaxy clusters in the tracer particle simulation. A few galaxy clusters have been examined. However, for the clarity of presentation, we only show results for the the first galaxy cluster in our sample which has $M_{\rm 500c} = 6.4\times 10^{14}\;\rm M_{\odot}$ and $r_{\rm 500c} = 1.3$ Mpc at $z=0$.
The selected galaxy cluster experienced an intense major merger epoch starting at around cosmic scale factor $a\approx 0.5$ ($t\approx 6$ Gyr), during which multiple small galaxy clusters of comparable sizes merged and formed a massive cluster (see Fig.\;\ref{fig:MAH}). The last small cluster joined the system at around $a=0.7$ and merged into the center of the system at about $a=0.75$. After this, the system stayed rather quiescent, had only a few minor mergers till the end of the simulation at $a=1$. We show images of the thermodynamic quantities of this selected galaxy cluster at various cosmic times in Fig.\;\ref{fig:image}, and the evolution of its density and entropy profiles in Fig.\;\ref{fig:profile_evo}. Since this single cluster has had various dynamical states during its growth history, the diversity of a sample of clusters induced by their different dynamical states is largely covered by studying the time evolution of this selected cluster. Indeed, we do not see qualitatively new features in other galaxy clusters in the simulation.

To study the growth of this galaxy cluster in its large scale environment, we select tracer particles within a large radius of $10 r_{\rm 500c}$ of this galaxy cluster at $z=0$. This results in a large number ($\gtrsim 10^6$) of tracer particles with a gas mass resolution of about $2.6\times 10^8 \;\rm{M}_{\odot}$. Then we trace back how the thermodynamical properties of these tracer particles evolved over space and time.

\section{Results}
\begin{figure}
\centering
    \includegraphics[width=.46\textwidth]{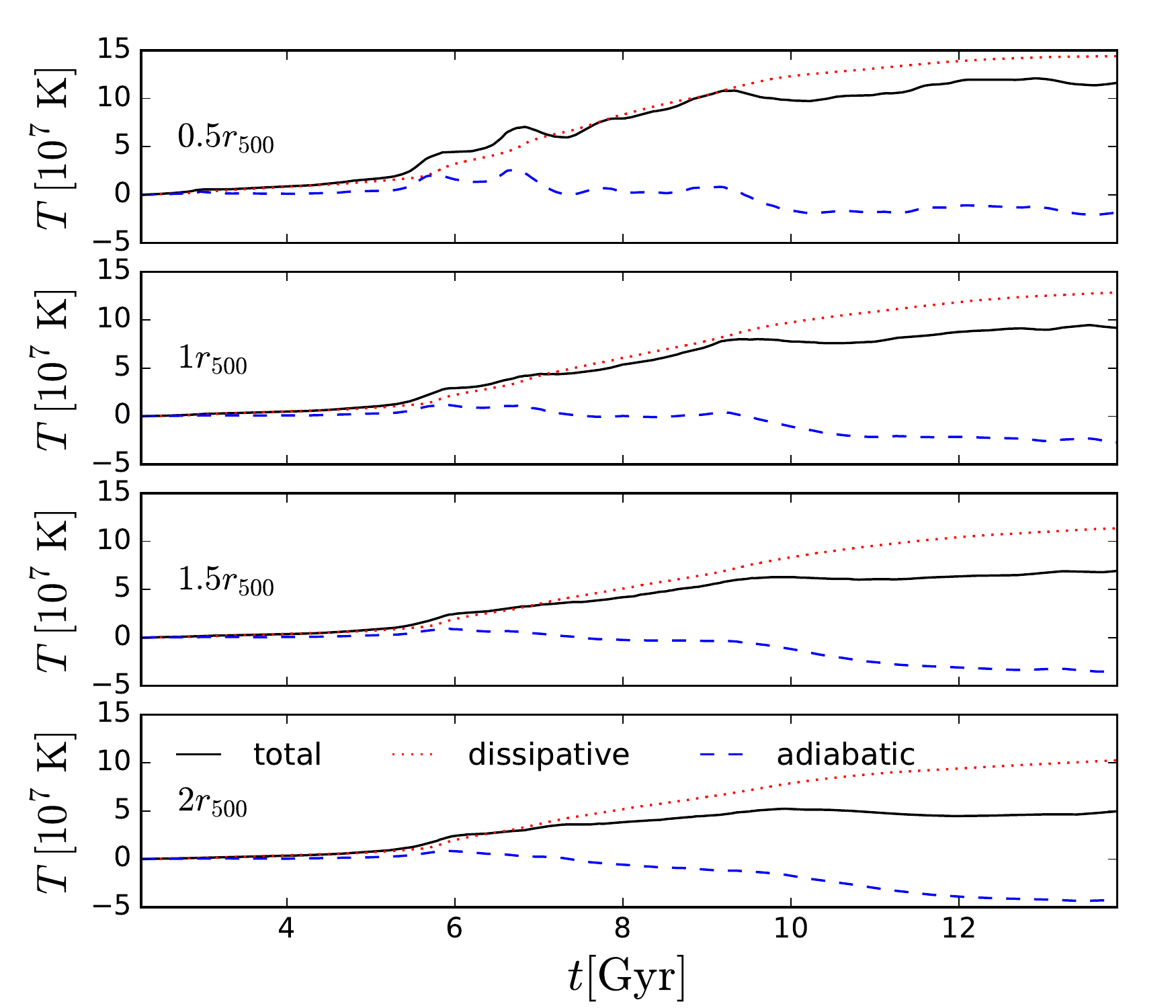}
  \caption{Thermal history of shells of ICM tracer particles selected around four different radii at $z=0$. Total temperature increase (black solid line) is divided into that from dissipative heating (red dotted line) and that from adiabatic heating (blue dashed line). The dissipative temperature increase is distributed over time, with no sign of accretion shock heating dominance which would mean concentrated heating at a certain cluster-entering epoch. At most times and radii, the temperature increase contributed by the adiabatic heating have negative values, suggesting a reduction of gas density with time on average, which also contradicts the smooth accretion model picture.}
\label{fig:heating_history}
\end{figure}

\begin{figure*}
\centering
\includegraphics[width=.75\textwidth]{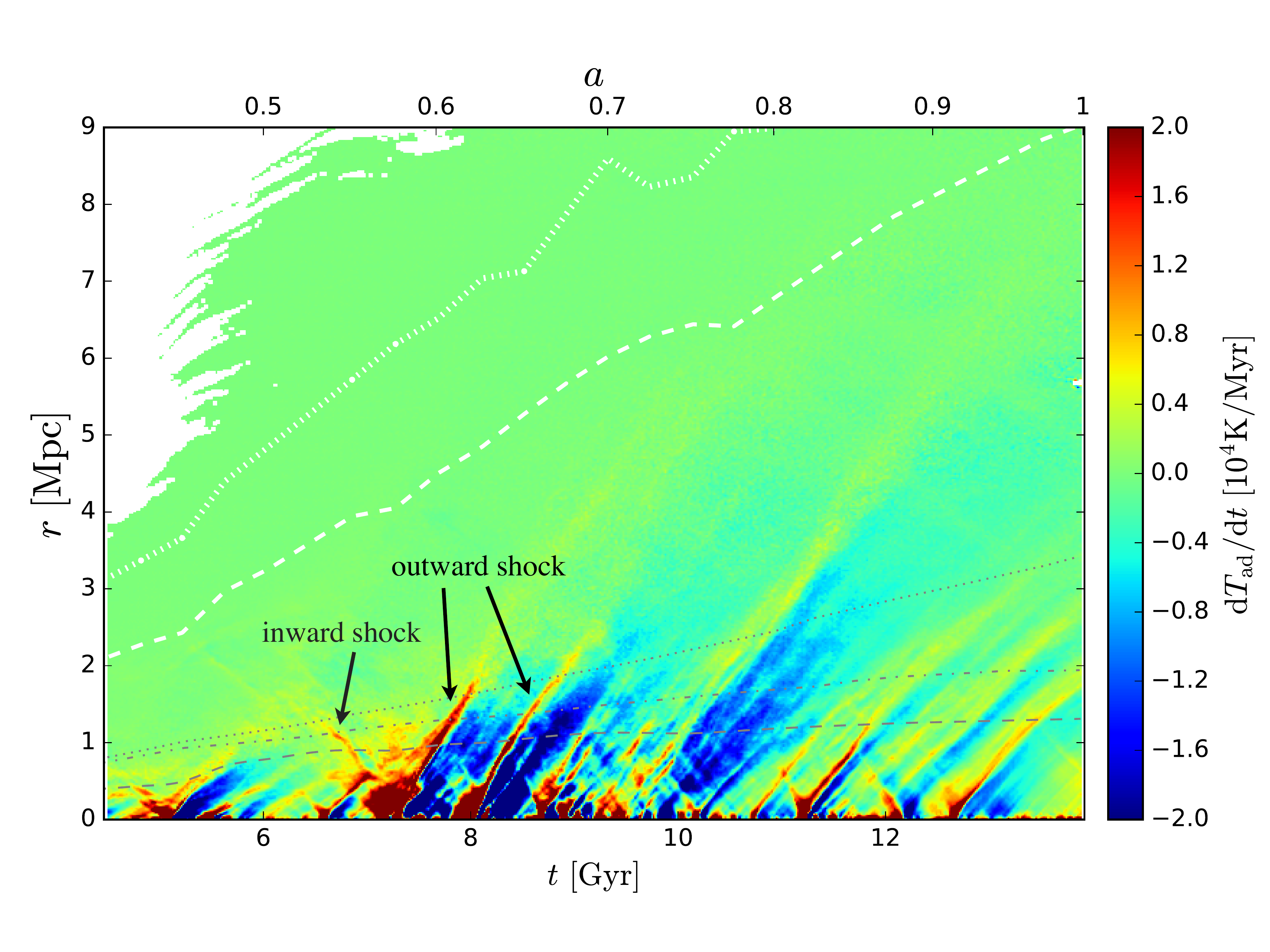}
  \caption{Distribution of adiabatic heating (and cooling) rate
  $\dd T_{\rm ad} / \dd t$ over radius $r$ and the cosmic time
  $t$ showing the spatial-temporal distribution of compression/rarefaction in the ICM of the selected galaxy cluster. Tracer particles with density above twice the median density in each $\dd r-\dd t$
  bin are ignored to show the adiabatic heating/cooling in the diffuse medium. The gray dashed, dash-dotted, and dotted lines mark the radii of $r_{\rm 500c}$, $r_{\rm 200c}$ and $r_{\rm 200m}$ respectively. The white dashed line indicates the volume-averaged accretion shock radius, whereas the white dotted line indicates the maximum accretion shock radius.  Outward propagating shocks leave prominent red/yellow colored tracks with positive slopes on the $r-t$ plane. They are followed by rarefaction waves indicated by the blue regions. For some mergers, an inward propagating shock that forms before the merging structure reaching its pericenter (red/yellow colored track with negative slopes on the $r-t$ plane) can also be identified.
 }
\label{fig:dTad_diffuse}
\end{figure*}

\begin{figure*}
\centering
    \includegraphics[width=.75\textwidth]{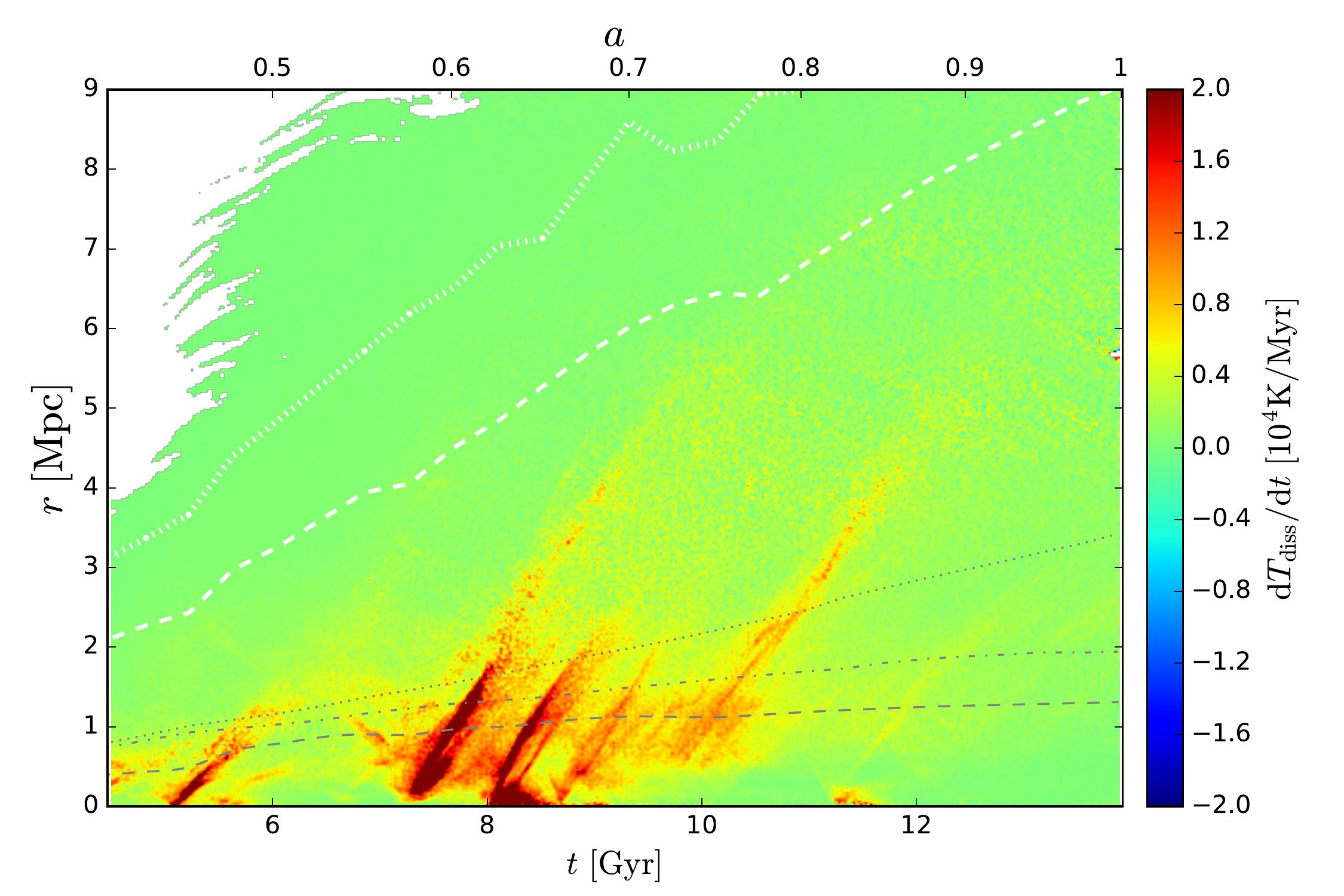}
  \caption{Same as Fig.\;\ref{fig:dTad_diffuse} but for the dissipative heating rate
  $\dd T_{\rm diss} / \dd t$. Compared to Fig.\;\ref{fig:dTad_diffuse}, only the few stronger shocks leave prominent tracks in this figure. There is dissipative heating not only at the compressive shock fronts but also in the rarefaction regions with negative $\dd \ln T_{\rm ad}$, suggesting the existence of heating mechanism other than direct heating at shock fronts.}
\label{fig:dTdiss_diffuse}
\end{figure*}

\subsection{Thermal history of the ICM}
\label{sec:thermal_history}
First of all, we examine the overall thermal history of the gas particles in the ICM: when did they get heated to their current temperature, and from what heating mode (dissipative or adiabatic)?

We expect the answer to depend on the distance of the gas particle to the cluster center, since a galaxy cluster builds up itself in an inside-out fashion. Fig.\;\ref{fig:heating_history} presents the average heating history of shells of tracer particles selected at various radii at $z=0$.
It shows that intracluster gas experiences continuous heating after accreting onto a galaxy cluster, rather than being heated dissipatively at a single epoch when it passes through the accretion shock and afterward experiences only adiabatic heating; namely, it does not follow the picture depicted by the smooth accretion models \citep{tozzi01, voit03}.

The heating mode that is responsible for the observed temperature of the ICM is the dissipative heating. The contribution from adiabatic heating is on average negative (i.e., cooling rather than heating) for all shells of tracer particles at most of the time. This reflects the fact that cosmic gas does not become denser when it enters a galaxy cluster. Rather, it preserves its high density at an early cosmic epoch better in the region of a galaxy cluster (c.f. top panels of Fig.\;\ref{fig:image}), but even there, the density slightly decreases with cosmic expansion as shown by the slight adiabatic cooling in the central regions (upper panels of Fig.\;\ref{fig:heating_history})\footnote{Radiative cooling, which can increase gas density in the core of a galaxy cluster, is neglected here.}.  The strong adiabatic cooling shown at the outer low-density regions (lower panels of Fig.\;\ref{fig:heating_history}) may be temporarily depending on the dynamical state of the cluster.

\subsection{Dynamical Heating of the ICM by Mergers}
We then ask the question where and when is the ICM getting heated and answer it by plotting the `dynamic profile' (i.e., the radial-temporal distribution of the heating rate) for the simulated cluster in Figs.\;\ref{fig:dTad_diffuse} and \ref{fig:dTdiss_diffuse}. Specifically, we bin the tracer particles in $\dd r - \dd t$ bins and plot the average compressive/dissipative heating rates of all tracer particles in each bin. By taking advantage of
the fine time resolution and a large number of tracer particles, we can take small bin sizes ($\dd r \approx 26$ kpc and $\dd t \approx$ 28 Myr) which allow us to obtain detailed maps of the heating history of the simulated galaxy cluster. We also separate the tracer particles into those tracing `diffuse' gas and `filamentary/clumpy' gas depending on whether their density is smaller or larger than two times the median density in the corresponding $\dd r - \dd t$ bin. In Figs.\;\ref{fig:dTad_diffuse} and \ref{fig:dTdiss_diffuse}, we plot only the contributions from the diffuse tracer particles to show the heating in the volume-filling diffuse ICM.

The location and time of intensive heating show patterns that are clearly related to mergers. In Fig.\;\ref{fig:dTad_diffuse}, one can easily identify more than ten heating (red and yellow) tracks, showing the prominent outward propagating (with positive slopes in the $r-t$ plane) merger shocks. These outgoing heating tracks are followed by cooling (blue) tracks, indicating rarefaction waves. The inward propagating (with negative slopes in the $r-t$ plane) shocks preceding the outward propagating ones are also evident for some merger events. They trace the shocks that formed before the pericentric passage of the merger.
In contrast to the large number of merger tracks observable in Fig.\;\ref{fig:dTad_diffuse}, the number of prominent merger tracks is much smaller in Fig.\;\ref{fig:dTdiss_diffuse}.
This suggests that mergers, even relative minor mergers, have a significant effect in adiabatically heating/cooling the ICM by varying the gas densities.
In comparison, only the major mergers are effective in causing the irreversible dissipative heating by varying gas entropies.
For our selected cluster, the two strongest mergers occurred during the epoch of major mergers. They can be identified by the two most intensive heating tracks in Fig.\;\ref{fig:dTdiss_diffuse}.

The contrast of the merger shock tracks in Figs.\;\ref{fig:dTad_diffuse} and \ref{fig:dTdiss_diffuse} reduces with distance. This is partially due to reduced shock heating at cluster outskirts where ICM temperature is lower, and partially due to broadened shock fronts at large distances due to anisotropic propagation and lower numerical resolution.

The accretion shock does not appear clearly in Figs.\;\ref{fig:dTad_diffuse} and \ref{fig:dTdiss_diffuse} as a prominent heating source even after taking account of the shock broadening effect. This carries the same message as Fig.\;\ref{fig:heating_history}, that the heating by mergers dominates over the heating at the accretion shock. This suggests that, in contrast to the `outside-in' heating depicted in the smooth accretion picture, the ICM is heated more in a `inside-out' fashion as kinetic energy from mergers drives thermodynamical heating starting from the cluster central regions to the outskirts. In other words, the heating from filamentary/clumpy accretion dominates over that from diffuse accretion although both accretion modes are important in building a cluster mass object, as the former deposits matter much deeper into the potential well and thus more energy can be converted into heat per particle. It would be interesting to compare the results with that for galaxy groups and galaxies, where radiative cooling further enhances the filamentary/clumpy mode of accretion.

\subsection{Radial-dependent heating mechanism}
\label{sec:heating_region}
\begin{figure}
\centering
    \includegraphics[width=.45\textwidth]{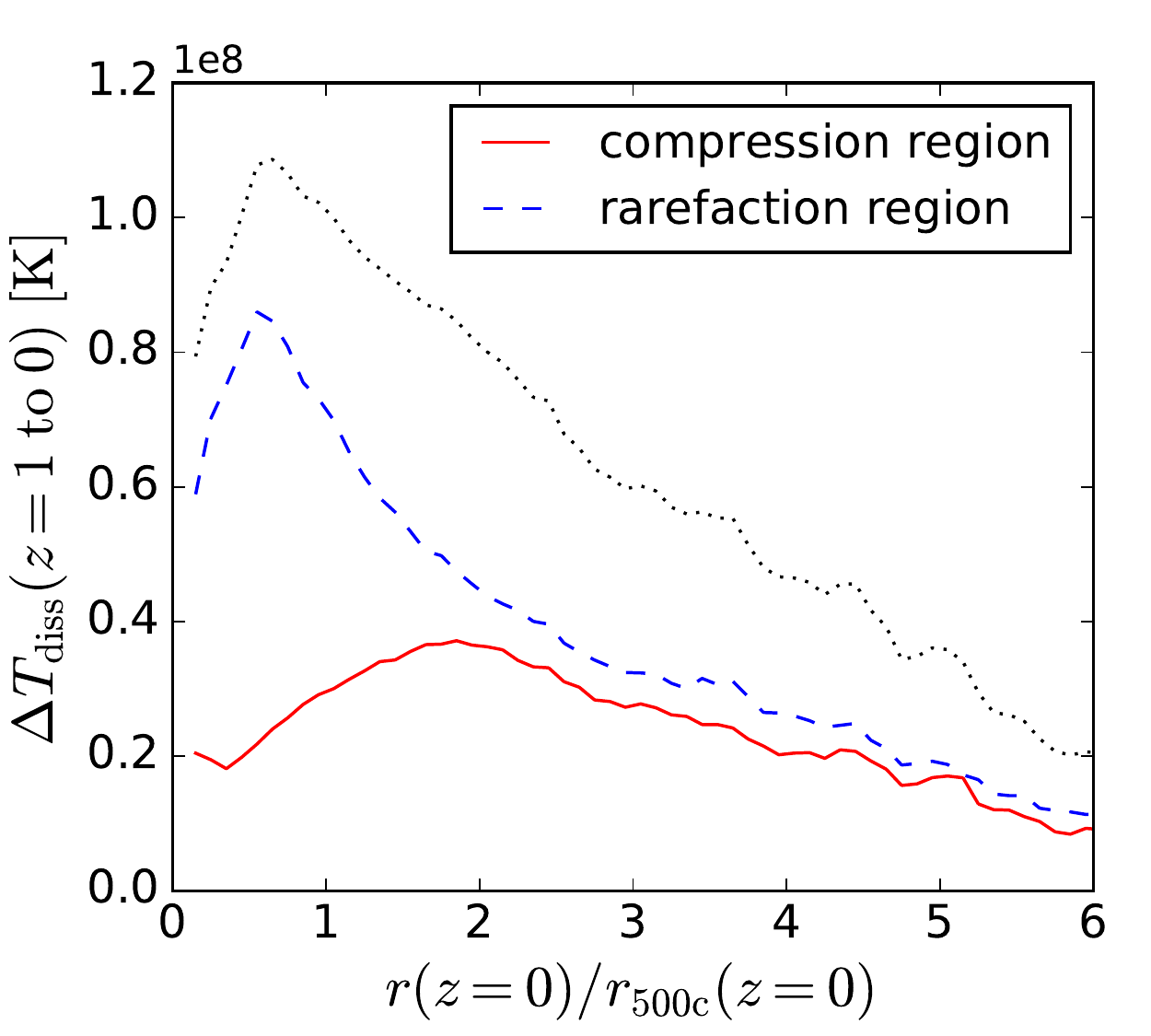}
  \caption{Dissipative heating $\Delta T_{\rm diss}$ for tracer particles between redshifts $z=1$ and $z=0$ as a function of the radius of the tracer particle at $z=0$ around the selected galaxy cluster. Total dissipative heating (black dotted line) is divided into that happened in the compression region (red solid line) and that happened in the rarefaction region (blue dashed line), with the former being contributed by both shock heating and turbulence heating, and the latter by turbulence heating alone.
  The dominance of $\Delta T_{\rm diss}$ in the rarefaction region over that in the compression region suggests the dominance of turbulence heating over shock heating, particularly at small radii. Here, compression / rarefaction is determined by the sign of a particle's density variation after ten snapshots ($\delta t \approx 250$ Myr) in order to suppress the noise from tiny density variations between subsequent snapshots. Tracer particles are averaged in radial bins of size $0.1 r_{\rm 500c}$ at $z=0$.}
\label{fig:heating_region}
\end{figure}

\begin{figure}
\centering
    \includegraphics[width=.47\textwidth]{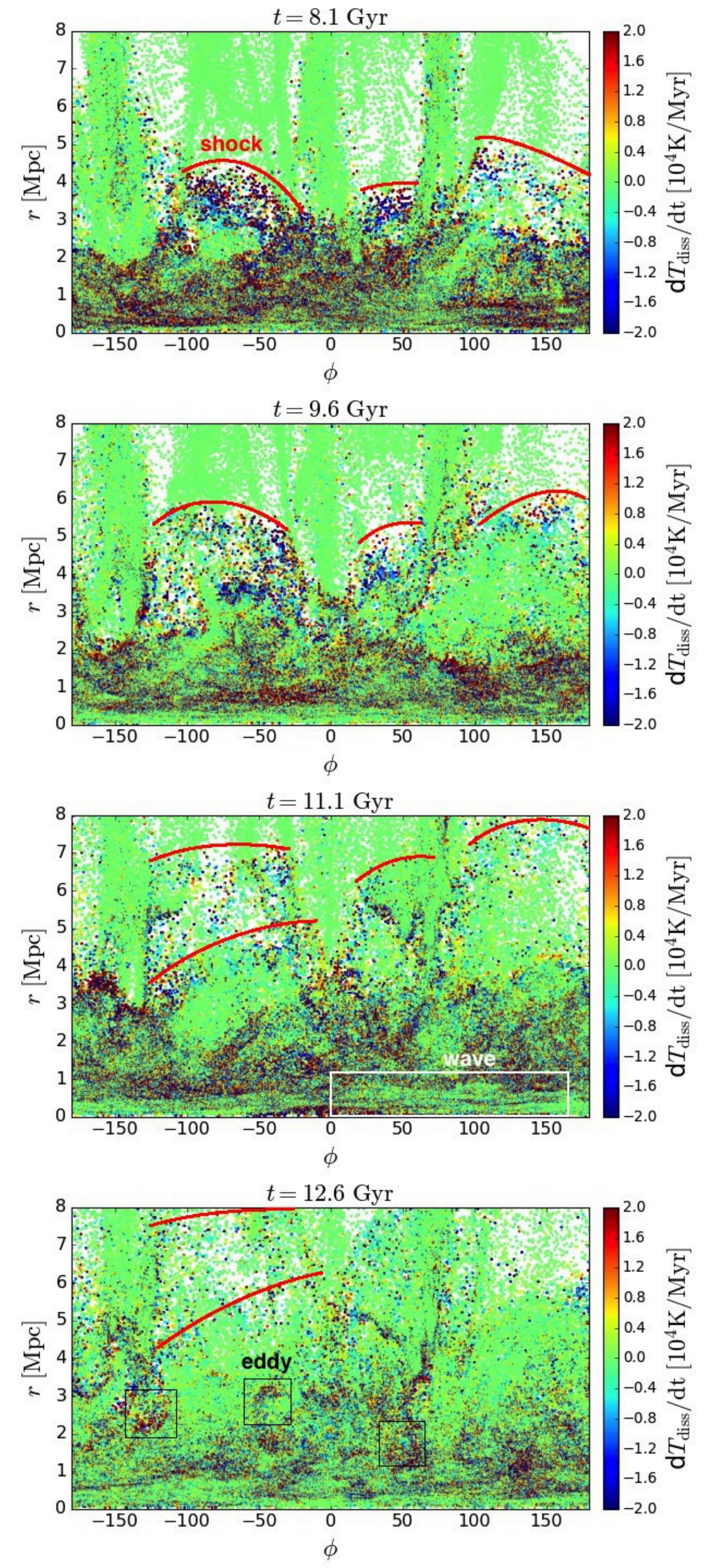}
  \caption{Evolution of the dissipative heating rate for tracer particles around the selected galaxy cluster in a thin $r-\phi$ slice with $9\uppi/20 < \theta < 11\uppi/20$ (same as the region presented in Fig.\;\ref{fig:image}). Outward propagating merger shock fronts can be seen at $-130<\phi<-20$, $10<\phi<50$, and $80 < \phi < 170$ degrees, as marked by the red curves. The outmost shock fronts have already joined with the accretion shock and mark the outer boundary of the ICM. In addition, eddy- and wave-like patterns (indicated with the black and white boxes, respectively) are present at small radii. The color-coding indicates the average dissipative heating rate in each $\dd r - \dd \phi$ bin. Regions not populated with any tracer particles are left white. Some regions have negative dissipative heating rates due to mixing effects (see Appendix.\;\ref{app:mixing_phys} and \ref{app:mixing_control}).}
\label{fig:dTdiss_rphi}
\end{figure}

In addition to direct heating at shock fronts, the ICM is heated also by the dissipation of kinetic energy contained in the gas motions.
The role of gas motions in heating the ICM has been studied in the cluster core regions that are affected by AGN feedback. Gas motions in these regions are expected to help balance radiative cooling and prevent a massive cooling flow in some cool core clusters \citep[e.g.,][]{zhuravleva14, zhuravleva18}.
The heating of merger-induced turbulent motions in the bulk of ICM, however, has not been studied systematically. The relative importance of such heating versus that of shock heating remains an open question.

Comparing the patterns of heating in Fig.\;\ref{fig:dTad_diffuse} and Fig.\;\ref{fig:dTdiss_diffuse}, one notices that there is heating associated with the merger but not directly contributed by the merger shock: the dissipative heating is much more extended in time than the adiabatic heating, and it extends well into
regions where $\dd T_{\rm ad} / \dd t < 0$ (Fig.\;\ref{fig:dTad_diffuse}); i.e., they occur in rarefaction regions where the gas density is decreasing. As dissipative heating at shock fronts occurs only in compression regions (since the shock front heats and compresses at the same time), the physical mechanism that generates dissipative heating in rarefaction regions must be subsonic. In our simulations, the only subsonic physical process that leads to heating is the dissipation of subsonic ICM motions. We refer to this as `turbulence dissipation' in general, since although a considerable fraction of subsonic kinetic energy of the ICM resides in bulk motions, it would dissipate into heat also through turbulence dissipation if the effective ICM viscosity is low. Thus, we take the dissipative heating in rarefaction regions as evidence of time-distributed turbulence dissipation of merger-generated ICM motions, which occurs in both the compression and rarefaction regions.


Therefore, the fraction of dissipative heating in compression versus rarefaction regions can serve as an indicator of the fraction of heating at shock fronts compared to turbulence dissipation heating. In order to estimate this fraction more quantitatively, we separate the dissipative temperature increment $\dd T_{\rm diss}$ for each tracer particle into that occurred in `compression region' ($\dd \rho_{\rm gas} > 0 $) and that in `rarefaction region' ($\dd \rho_{\rm gas} < 0 $), and compute the average contribution from both regions for tracer particles in radial bins of size $0.1 r_{\rm 500c}$ at $z=0$.
The result is presented in Fig.\;\ref{fig:heating_region} for the temperature increment $\Delta T_{\rm diss}$ between $z=1$ and $z=0$.

As shown by Fig.\;\ref{fig:heating_region}, heating in rarefaction regions, and thus turbulence heating, dominates the total amount of dissipative heating that has occurred between redshift one and zero at inner radii $r \lesssim r_{\rm 500c}$.\footnote{Mixing among neighboring gas particles affects the computation of the relative heating in the compression and rarefaction regions, as will be discussed later in this section and the appendices. We, therefore, concentrate only on the most reliable features in Fig.\;\ref{fig:heating_region}. For example, we consider the heating in the rarefaction regions as dominating only where it is significantly (more than two times) greater over the heating in the compression region. This results in a conservative radial limit of this domination, $r \lesssim r_{\rm 500c}$.} This is consistent with the low-Mach number ($M\sim 2-4$) typical of merger shocks in this intracluster region, which implies a weak dissipative heating efficiency \citep{ryu03,kang07} and that most of the coherent kinetic energy associated with mergers is transferred into random gas motions in the ICM \citep{shi14}. In comparison, \citet{miniati15} found that the fractional heating contributed by numerically resolved turbulence dissipation is about one third in the inner regions ($r<r_{\rm vir}/3$) in their simulated galaxy cluster.

Outside $r_{500\rm c}$, turbulence heating decreases with radius (Fig.\;\ref{fig:heating_region}), which appears to be the major cause of a reduced temperature at cluster outskirts compared to the inner regions. This is at least partially due to the longer turbulence dissipation time of ICM gas motions at large radii, as suggested in \citet{shi14} and shown by a multi-scale analysis of simulated galaxy clusters by \citet{shi18}. Physically, a longer turbulence dissipation time at large radii in galaxy clusters is likely the result of weaker density stratification and a subsequent longer buoyancy time there \citep{shi19}. This radial dependent turbulence dissipation time and the subsequent radial dependent heating could also explain the deviation of the gas temperature profile in the cluster outskirts from self-similar evolution \citep[see][and references therein]{avestruz16}. In a broader picture, this provides a detailed example of how mass accretion influences the self-similarity of gas profiles in galaxy clusters \citep{lau15}.

The ratio of dissipative heating happened in the compression region (red solid line) versus that in the rarefaction region (blue dashed line) increases with radius. As shock heating contributes only to the red solid line and turbulence heating contributes to both lines, the increasing ratio suggests a stronger role of shock heating at larger radii. Apart from the less efficient turbulence dissipation at large radii as described above, this is also contributed by the larger Mach number of a typical merger shock at large radii (see Sect.\;\ref{sec:Mach}) which leads to stronger shock heating there.

When interpreting the results related to Fig.\;\ref{fig:heating_region}, one must be aware of the possible contamination from mixing between gas elements of different densities.
Isobaric mixing between gas elements of different densities does not lead to heating on average (see Appendix.\;\ref{app:mixing_phys}). However, it introduces an anti-correlation between $\dd K$ and $\dd \rho_{\rm gas}$ which mimics heating in rarefaction regions. Since mixing occurs predominantly at boundaries between diffuse ICM and dense structures like filaments and clumps, we limit its effect by restricting our analysis to diffuse regions. Additionally, we apply either spatial or temporal binning in our analysis in this paper to reduce mixing effects in diffuse regions (see Appendix.\;\ref{app:mixing_control}). Specifically, for the analysis presented in Fig.\;\ref{fig:heating_region}, we have used a time step that is ten times the time resolution of the tracer particle simulation outputs when identifying the sign of $\dd \rho_{\rm gas}$. The choice of the time step affects the quantitative result. We discuss the choice of binning in  Appendix.\;\ref{app:mixing_control} and show that our main results here, i.e., the increasing importance of shock heating with radius and the dominance of turbulence heating at inner regions, are robust against it.

In the spatial distribution map of the dissipative heating rate (Fig.\;\ref{fig:dTdiss_rphi}), there exists patterns directly indicative of the heating mechanism. In a few directions ($-130<\phi<-20$, $10<\phi<50$, and $80 < \phi < 170$ degrees) in Fig.\;\ref{fig:dTdiss_rphi}, one can clearly see dissipation at merger shock fronts propagating outward with time, reaching large radii. In some directions, multiple layers of merger shock waves can be identified. We have marked the locations of the most prominent shock fronts with red curves. At small radii ($r<4 \;\rm Mpc$), there are eddy-like patterns, for example the ones indicated with black boxes in the bottom panel, suggesting dissipation of turbulent gas motions. In the innermost regions ($r\lesssim 1 \;\rm Mpc$, still much larger than the $r\sim 100 \;\rm kpc$ core region where radiative cooling could be important) in each panel, one can see stratified, wave-like patterns (most clearly shown in the region marked by the white box in the third panel). They are indicative of dissipation of gravity waves, which in this paper we classify as part of the `turbulent motions'. The fact that the merger shocks appear across a large range of radius and the turbulent motions are more prominent at relatively small radii is consistent with the quantitative results shown in Fig.\;\ref{fig:heating_region}.

Further support of the dominance of turbulence heating comes from the velocity dispersion of the ICM. Turbulence dissipation rate is usually estimated with turbulence velocity dispersion $\sigma$ and the scale $\ell$ where its energy peaks as $\sim \sigma^3 / \ell$, which describes that the specific kinetic energy of the turbulence $\sim \sigma^2$ dissipates in a turn-over time scale $\sim \ell/\sigma$ of the energy-containing eddies. Using the radial velocity dispersion measured in radial bins of the original Eulerian simulation as an approximation for $\sigma$, and a fixed scale of $\ell =200\ \rm kpc$ as a typical scale of energy-containing eddies in the ICM (cf. Fig.\; of \citet{shi18}), we derive a rough estimate of the evolution of turbulence heating rate profile (bottom panel of Fig.\;\ref{fig:profile_evo}). Thus estimated heating rates are a few times more than that shown in Fig.\;\ref{fig:dTdiss_diffuse}, which is likely due to some bulk motions which have not yet entered turbulence cascade (but will at a later time) and have contributed to the measured velocity dispersion. Nevertheless, this estimation confirms that there is sufficient kinetic energy contained in ICM motions to be responsible for the measured rates of dissipative heating. Moreover, the radial and temporal dependencies of the estimated turbulence heating rate as presented in the bottom panel of Fig.\;\ref{fig:profile_evo} are consistent with the measured $\dd T_{\rm diss} / \dd t$ shown in Fig.\;\ref{fig:dTdiss_diffuse}, both peaking at passages of the strong merger shocks.

\subsection{Properties of merger shocks}
\label{sec:merger_shock}

\subsubsection{Propagation of shock fronts}
Another aspect of the merger shocks revealed by the dynamic profile shown in Figs.\;\ref{fig:dTad_diffuse} and \ref{fig:dTdiss_diffuse} is their trajectories.

First of all, it is striking to notice that almost all merger shocks with clear trajectories in Fig.\;\ref{fig:dTdiss_diffuse}, both major and minor ones, propagate far beyond $r_{\rm 500c}$ and thus belong to the `runaway merger shocks' \citep{zhangcy19b}. Some major merger shocks even extend to the cluster periphery, reaching the location of the accretion shock (white dashed line in Fig.\;\ref{fig:dTdiss_diffuse}).

Secondly, the tracks of the outward propagating merger shocks, especially those of the major merger shocks, are rather straight in the $r-t$ plane. This suggests no significant change in the shock propagation velocity in the diffuse regions over a large range of distance $0<r\lesssim 5$ Mpc, despite a significant change in the sound speed in this radial range. 

The spatial distribution map of the dissipative heating rate (Fig.\;\ref{fig:dTdiss_rphi}) shows an additional feature of the merger shocks: its propagation is highly anisotropic at large radii. The shock fronts propagate the fastest in directions with low gas densities and are slow, sometimes even quenched, in directions with high-density filaments (see also Fig.\;\ref{fig:image}).

When freely propagating in directions away from the filaments,
the strength of merger shocks depends mainly on the density slope of the ICM, which is relatively flat ($\rm {d} \ln \rho_{\rm gas} / \rm{d} \ln r \approx -1$) in the central region and progressively steeper in the outskirts. Once a shock reaches a region with a sufficiently steep density slope of $\rm {d} \ln \rho_{\rm gas} / \rm{d} \ln r \lesssim -3$, the shock does not get attenuated with time and thus can be `long-lived' and propagate to very large distances \citep{zhangcy19b}.
On the other hand, when the core of a merging subcluster survives the merger event, it can travel at supersonic speed for some distance, push the merger shock and accelerate it even in regions with flat density slopes \citep[see Fig.\;3 of][]{zhangcy19a}. The evolution of a merger shock as it propagates is usually determined by the combination of these two effects.
For the major merger events of the galaxy cluster presented in this paper, the cores of the merging galaxy clusters are destroyed during the merger, but some gas particles are accelerated to high speed by the merger. This ejecta moves outward supersonically with little resistance from the surrounding diffuse gas in directions away from the filaments, reminiscent of the free expansion phase of supernova remnant.
For some major mergers, the ejecta carries a sufficient amount of momentum so that it is not stopped by the diffuse ICM. Instead, the ejecta propagates to the cluster periphery and what finally stops it is the intergalactic gas flow accreting onto the galaxy cluster. Thus, the merger shock driven by this ejecta moves outward all the way to the accretion shock location, where it joins with the accretion shock and becomes the outer boundary of the heated ICM gas. The shock that formed out of the interaction between the merger shock and the accretion shock, which a recent study by \citet{zhangcy20} refers to as the Merger-accelerated Accretion shock (MA-shock), is long-living and can propagate to very large distances. The outmost shocks in Figs.\;\ref{fig:image} and \ref{fig:dTdiss_rphi} are actually MA-shocks.

\subsubsection{Dissipation vs compression at merger shocks}
\label{sec:Mach}
\begin{figure}
\centering
    \includegraphics[width=.48\textwidth]{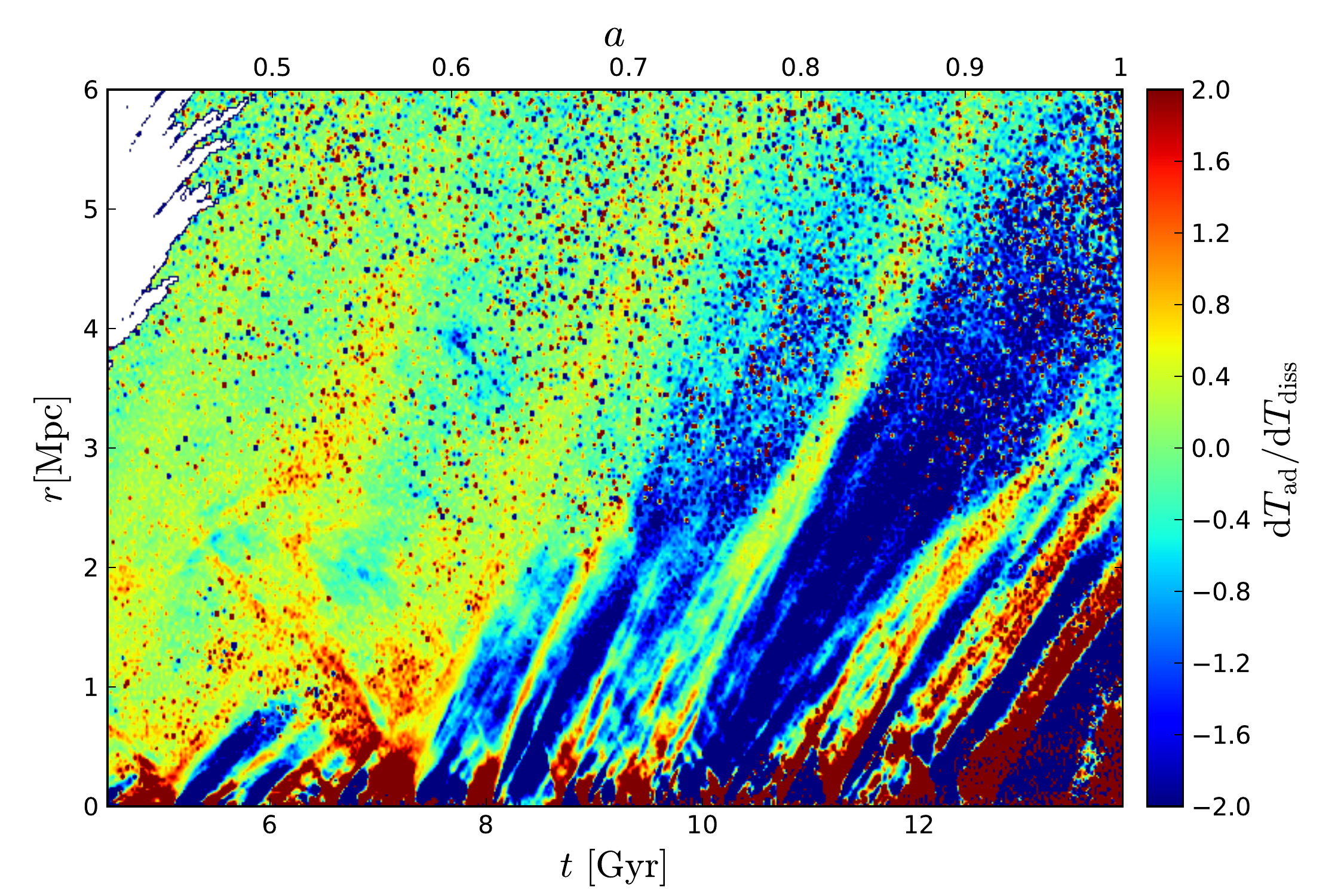}
  \caption{Ratio of adiabatic and dissipative heating rates plotted in Figs.\ref{fig:dTad_diffuse} and \ref{fig:dTdiss_diffuse}. At the tracks of merger shocks, this ratio can act as an indicator for the Mach number of the shocks, with a lower but positive $\dd \ln T_{\rm ad} / \dd \ln T_{\rm diss}$, indicating a higher Mach number (see Fig.\;\ref{fig:Mach}).  }
\label{fig:ad_diss_ratio_diffuse}
\end{figure}

\begin{figure}
\centering
    \includegraphics[width=.37\textwidth]{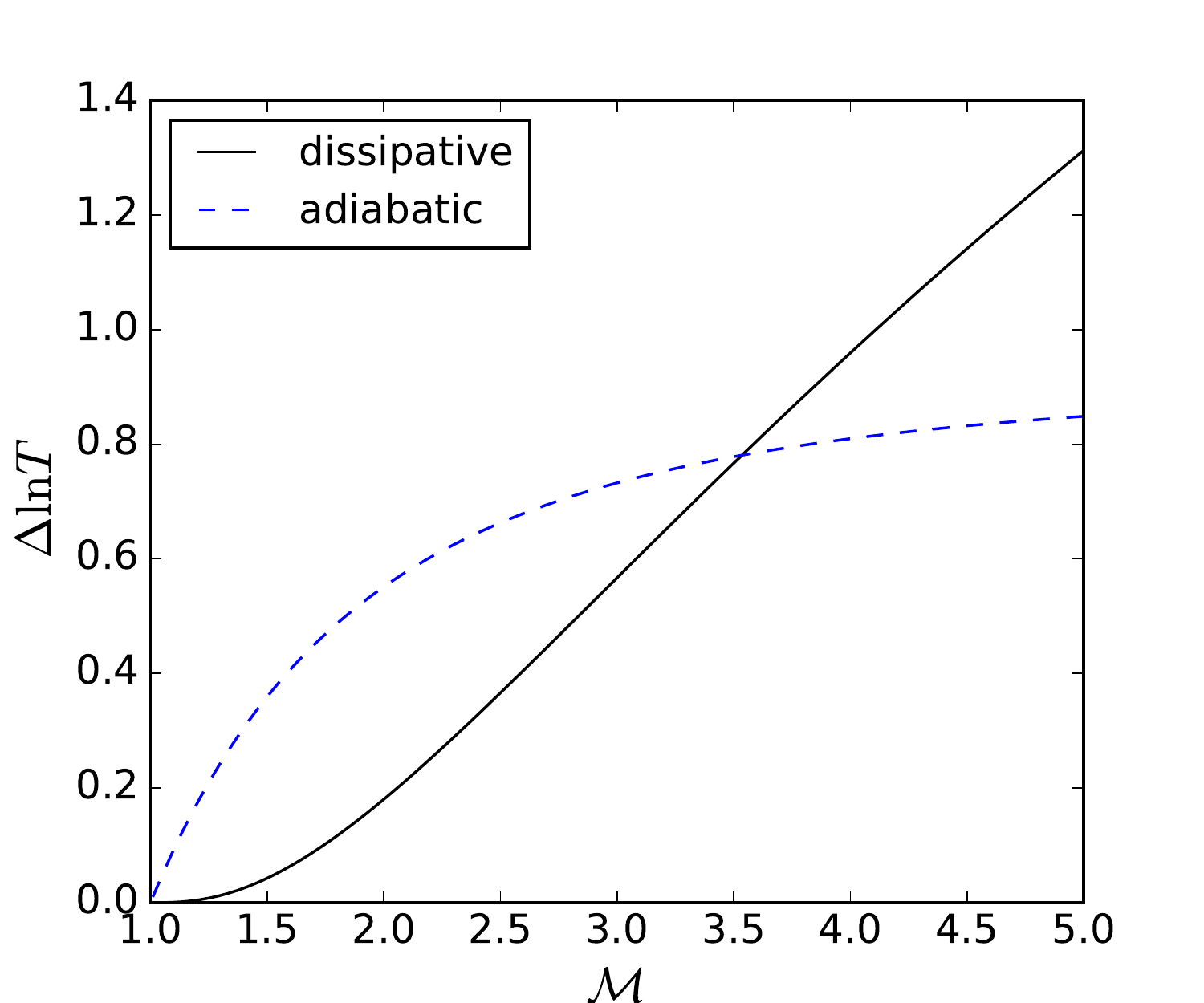}
  \caption{Dissipative versus adiabatic heating at shocks as a function of the Mach number $\mathcal{M}$ derived from the theoretical Rankine-Hugoniot shock jump conditions
   for gas with adiabatic index $\gamma=5/3$ (see Eqs.\ref{eq:shockheating} and \ref{eq:jump_condition}).}
\label{fig:Mach}
\end{figure}

The ratio of adiabatic and dissipative heating rates at the merger shock trajectories (Fig.\;\ref{fig:ad_diss_ratio_diffuse}) gives us information about the Mach number of the shock, as the ability of merger shocks in heating and compressing the gas is determined by the latter. Specifically, consider the fractional temperature increase $\Delta \ln T$ of the gas across the shock front. The dissipative heating and adiabatic heating contributions are
\eqs{
\label{eq:shockheating}
& (\Delta \ln T)_{\rm diss} = \Delta \ln K  =
\Delta \ln P  - \gamma \Delta \ln \rho_{\rm gas} = \ln \br{\frac{P_2}{P_1}} - \gamma \ln
\br{\frac{\rho_{\rm gas, 2}}{\rho_{\rm gas, 1}}} \,,\\
&(\Delta \ln T)_{\rm ad} =  (\gamma - 1) \Delta
\ln \rho  = (\gamma - 1) \ln \br{\frac{\rho_{\rm gas, 2}}{\rho_{\rm gas, 1}}} \,,
}
where the ratios of densities and pressures across the shock can be derived from
the Rankine-Hugoniot shock jump conditions as 
\eqs{
\label{eq:jump_condition}
\frac{\rho_{\rm gas, 2}}{\rho_{\rm gas, 1}} & = \frac{(\gamma+1) M_1^2}{(\gamma-1)M_1^2+2} \,,\\
\frac{P_2}{P_1} & = \frac{2\gamma M_1^2 - (\gamma-1)}{\gamma+1} \,.
}

We plot $(\Delta \ln T)_{\rm diss}$ and $(\Delta \ln T)_{\rm ad}$ as functions
of the shock Mach number in Fig.\;\ref{fig:Mach} for $\gamma=5/3$. Apparently, dissipative
heating is a stronger function of the Mach number, whereas adiabatic heating
saturates at high Mach number. The two effects contribute an equal amount of gas heating across the shock front at $\mathcal{M} \approx 3.5$, above which dissipative heating dominates, and below which adiabatic heating dominates. The ratio of the two heating terms,
${\Delta T_{\rm ad}}/{\Delta T_{\rm diss}}$, decreases monotonically with the Mach number. It approaches infinity at $\mathcal{M}\to 1$ and zero at $\mathcal{M}\to \infty$.

From Fig.\;\ref{fig:ad_diss_ratio_diffuse}, we can see that the $\dd \ln T_{\rm ad} / \dd \ln T_{\rm diss}$ ratios for the two major outward propagating shocks are below unity for $r>1$~Mpc. Comparing this to the theoretical estimation above suggests that they have $\mathcal{M} > 3.5$ there. On Fig.\;\ref{fig:image} we indeed observe high temperature jumps assiciated with the shocks in the outer regions. For example, around the northwest direction ($\phi = 3\pi/4$) of the ICM at $t=11.1\ \rm Gyr$ (3rd column of Fig.\;\ref{fig:image}), the temperature jump $T_2 / T_1$ is $\lesssim 5$ for the merger shock at $r\gtrsim 4\ \rm Mpc$, corresponding to a Mach number of $\sim 3.5$. The outmost MA-shock around 8 Mpc is associated with a more dramatic temperature jump of more than two orders of magnitude, suggesting a Mach number much greater than 10.
The later outward propagating shocks starting at $t>10$ Gyr, in comparison, have larger $\dd \ln T_{\rm ad} / \dd \ln T_{\rm diss}$ ratios which remain greater than unity out to $2-3$ Mpc. This is consistent with them being low Mach number weak shocks induced by minor mergers.

\section{Conclusions}
We have performed a detailed study of ICM heating by analyzing a tracer particle re-simulation of the non-radiative Omega500 simulation. The high time resolution and the Lagrangian nature of the tracer particles allowed us to decompose the heating of ICM into dissipative and adiabatic heating, depending on whether it is associated with entropy increase or not. Combined with the high spatial resolution provided by a large number of tracer particles, we could identify the location, time and mode of heating throughout the mass assembly history of a galaxy cluster. In particular, we have studied the relative contribution of dissipative vs adiabatic heating, and that of shock heating vs turbulence dissipation.

Our main result is that the numerous merger events are the prime heating source of the ICM, as visualised by Figs.\;\ref{fig:dTad_diffuse} and \ref{fig:dTdiss_diffuse}. Merger events heat the ICM both directly at merger shock fronts and in a delayed fashion by the dissipation of merger-induced turbulence. The latter dominates the heating in the inner cluster regions of $r<r_{\rm 500c}$, while the fractional contribution of merger shock heating increases with radius (Fig.\;\ref{fig:heating_region}). Outside $r_{\rm 500c}$, turbulence heating decreases with radius, leading to lower temperatures at cluster outskirts compared to the inner regions. Overall, the heating of the ICM is contributed by dissipative processes distributed over time and space, and more in an `inside-out' fashion as initiated by the mergers rather than in an  `outside-in' fashion by the accretion shock, a picture very different from that described in the smooth accretion model.

Additionally, our analysis has revealed some under-appreciated and new aspects of ICM merger shocks. In particular, the dynamic profile  ($r-t$ distribution) of the adiabatic and dissipative heating rates in the ICM (Figs.\;\ref{fig:dTad_diffuse} and \ref{fig:dTdiss_diffuse}) prove to be an excellent tracer of merger shocks. While the adiabatic heating rate distribution (Fig.\;\ref{fig:dTad_diffuse}) reflects the propagation of shock compression and subsequent rarefaction from both major and minor mergers, the dissipative heating rate distribution (Fig.\;\ref{fig:dTdiss_diffuse}) reflects the irreversible heating at and following the major merger shocks. A combination of them reveals both the shock trajectories and their Mach numbers (Fig.\;\ref{fig:ad_diss_ratio_diffuse}).
Remarkably, merger shocks can frequently run away (i.e., propagate to very large distances much greater than $r_{\rm 500c}$ in directions away from filaments), approach and join the accretion shock in the cluster outskirts.

Our results emphasize the role of mergers, shocks, and turbulence in the growth of galaxy clusters, and highlight a physical link between the non-thermal and thermal properties of the ICM. In the near future, XRISM and Athena will provide direct measurements of gas motions, complementing detailed ALMA, Chandra, LOFAR, XMM-Newton observations of thermal and non-thermal processes in the ICM. The proposed SKA, Lynx, and Cosmic Web Explorer missions have the potential to further extend these measurements to the outer parts of galaxy clusters. Our findings provide a theoretical basis for interpreting future X-ray and SZ observations from these probes, and will advance applications of X-ray and microwave measurements of galaxy clusters for cosmology.

\section*{Acknowledgements}
We are grateful to Congyao Zhang, Erwin Lau, Eugene Churazov, Irina Zhuravleva, and the anonymous referee for helpful discussions and comments on the draft of this paper. X.S. acknowledges support by NSFC Grant No. 11973036. D.N. acknowledges Yale University for granting a triennial leave and the Max-Planck-Institut f\"{u}r Astrophysik for hospitality where this work was performed. The simulations were performed on the Omega HPC cluster at Yale. This work is supported in part by NSF AST-1412768 and the facilities and staff of the Yale Center for Research Computing.

\bibliographystyle{mnras}
\bibliography{bibliography}

\appendix

\section{Accuracy of tracer particle simulation}
\label{app:tracer}

\begin{figure}
\centering
    \includegraphics[width=.37\textwidth]{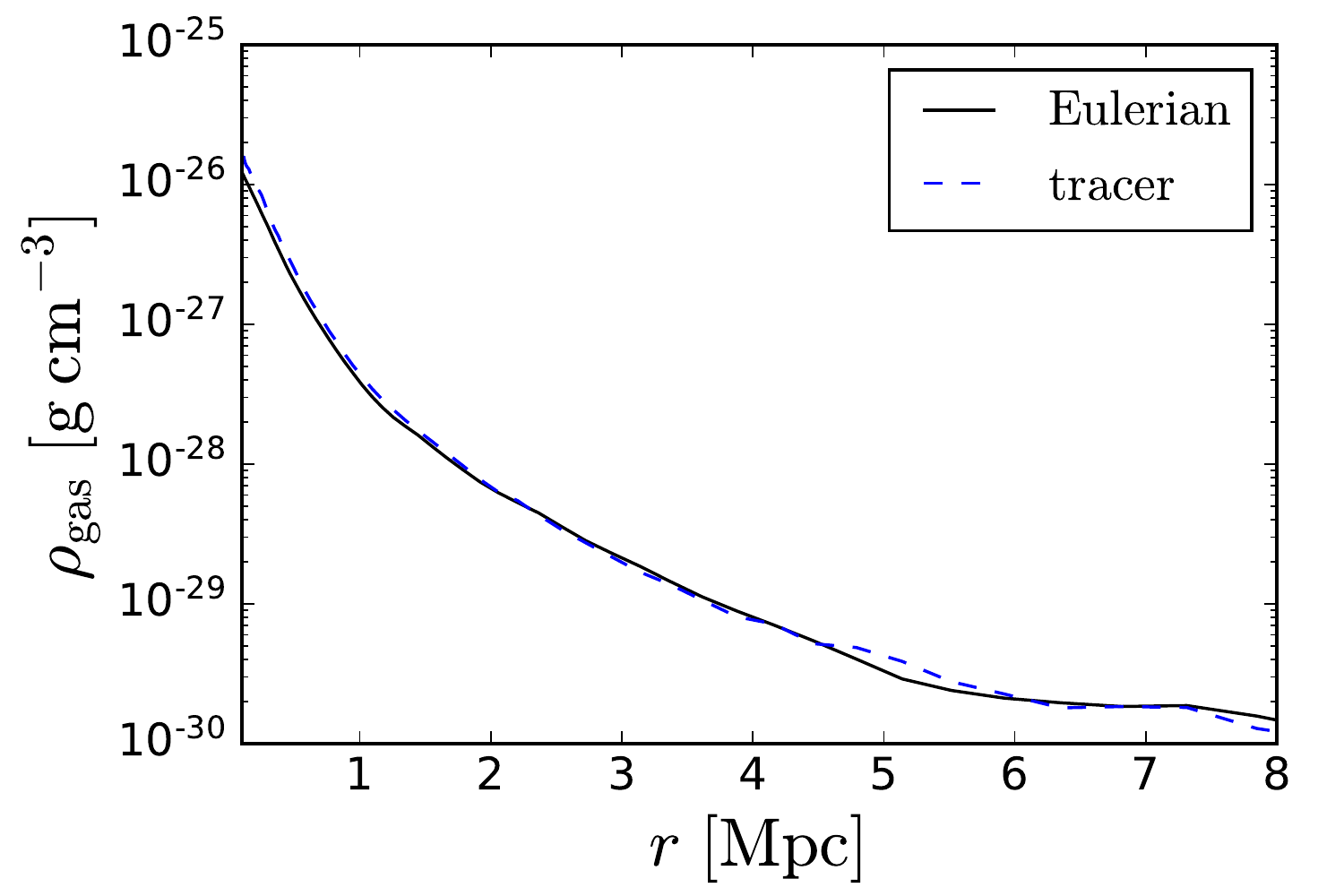}
  \caption{Comparison of the gas density profiles in the original Eulerian simulation (black solid line) and that computed using the number density of the tracer particles (blue dashed line) at redshft zero, showing good consistency between the two. Different from Fig.\;\ref{fig:profile_evo}, no clump removal has been applied to the profiles here to enable a fair comparison. }
  \label{fig:tracer_densityprof}
  \end{figure}

\begin{figure}
\centering
    \includegraphics[width=.42\textwidth]{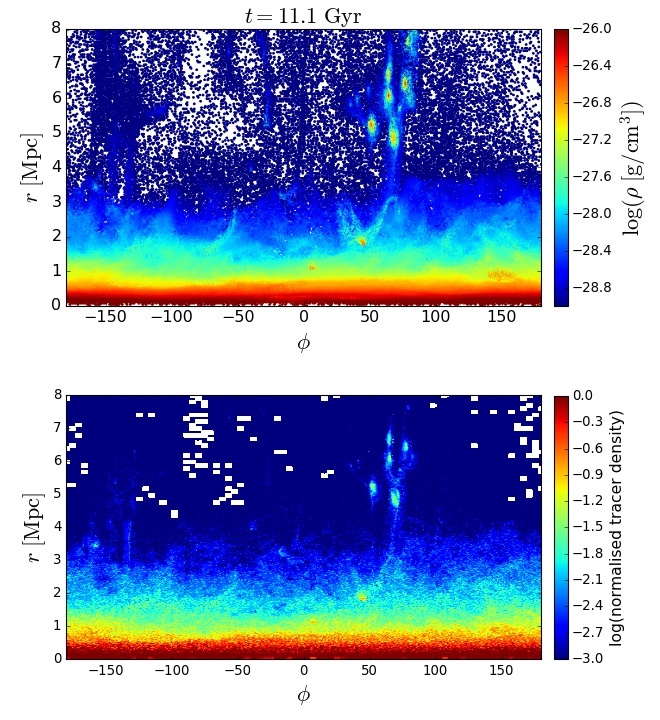}
  \caption{Comparison of the gas density gas density of Eulerian cells carried by the tracer particles (upper panel) and the number density of the tracer particles (lower panel) in the same $r-\phi$ slice of the galaxy cluster as presented in Fig.\;\ref{fig:dTdiss_rphi}. The number density of the tracer particles is computed from the number of tracer particles in finite $r-\phi$ bins of $20 \;\rm{kpc}$ by $0.9$ degrees. The overall consistency of the tracer particle number density with the underlying gas density distribution demonstrates the fidelity of the tracer particle re-simulation.}
\label{fig:tracer_density}
\end{figure}

Tracer particle simulation is a re-simulation of a Eulerian simulation using Lagrangian particles. It has been applied to a wide range of astrophysical topics, for example, thermodynamical history of gas elements \citep{vazza11b, genel13,cadiou19}, acceleration of cosmic rays \citep{wittor17a} and turbulent mixing \citep{federrath08, vazza10a}. It inherits the advantages of a Eulerian simulation, for example, the better characterization of fluid instabilities and thus the turbulent motions comparing to a Lagrangian simulation of the same resolution. However, accurately tracing mass flow using passive tracer particles is technically non-trivial. Due to the finite spatial and temporal resolution of the velocity field used to advect the tracer particles, small errors occur in the locations of the tracer particles. As a result, the density of the tracer particles may deviate from the underlying gas density in the Eulerian simulation, as has been pointed out in \citet{genel13,cadiou19}. Numerical resolution plays a key role in determining the degree of this deviation. In low-resolution cosmological simulations, this deviation could be an order-of-magnitude effect for low-mass halos, but when the resolution increases it becomes much milder (Fig.\;13 of \citealt{genel13}). Besides, radiative cooling and the consequent converging flows usually lead to large local deviations. This effect is not severe for our simulation given its high spatial and temporal resolutions and its non-radiative nature. We show a comparison of the density profiles in Fig.\;\ref{fig:tracer_densityprof}, where we observe an overdensity of the tracer particles in the central regions, a caveat which has been reported in the literature cited above. However, in our case the deviation is very mild. Actually, the consistency between the gas density profiles in our simulations is better than that observed in \citet{genel13} even when compared to the stacked profiles of their `high mass' halos which are still less massive and less well-resolved than our clusters.
We also present a comparison of the density distributions in Fig.\;\ref{fig:tracer_density}, which shows that most structures in the ICM are indeed well-captured by the tracer particles.

The finite accuracy of the tracer particle advection scheme has another trickier effect that is more relevant for this study: a spurious mixing of the tracer particles which is an error in matching gas elements in different snapshots. In other words, the tracer particles are more diffusive compared to the particles in a true Lagrangian simulation. We discuss the control of mixing effects including this spurious mixing in Appendix\;\ref{app:mixing_control}.

\section{Isobaric mixing}
\label{app:mixing_phys}

Here we consider the mixing of gas elements under pressure equilibrium. As is well known, mixing increases entropy, but as we will demonstrate below, there is no change in average $T_{\rm diss}$ or $T_{\rm ad}$.

For the clarity of demonstration, we consider two equal mass gas elements. They have different densities and temperatures before mixing and the same after mixing. For each gas element,
\eqs{
\Delta T_{\rm diss} & =  \int_{\rho_{\rm i}} ^{\rho_{\rm f}}  T(\rho) \dd \ln K(\rho) \\
& =  - \gamma T_{\rm i} \int_{\rho_{\rm i}} ^{\rho_{\rm f}} \frac{\rho_{\rm i}}{\rho} \dd \ln (\rho / \rho_{\rm i}) \\
& = \gamma T_{\rm i} \br{\frac{\rho_{\rm i}}{\rho_{\rm f}} - 1}\\
& = \gamma (T_{\rm f} - T_{\rm i})\,,
}
and similarly,
\eqs{
\Delta T_{\rm ad} & =  (\gamma-1) \int_{\rho_{\rm i}} ^{\rho_{\rm f}}  T(\rho) \dd \ln \rho \\
& =  (\gamma-1) T_{\rm i} \int_{\rho_{\rm i}} ^{\rho_{\rm f}} \frac{\rho_{\rm i}}{\rho} \dd \ln (\rho / \rho_{\rm i}) \\
& = -(\gamma-1) (T_{\rm f} - T_{\rm i})\,,
}
where `i' and `f' indicate the states before and after mixing, respectively.

Since the total internal energy is conserved throughout the mixing, $2 T_{\rm f} =  T_{\rm i, 1} + T_{\rm i, 2}$, the change in $T_{\rm diss}$ from the two elements have opposite signs and cancel each other, 
\eqs{
& \Delta T_{\rm diss, 1} + \Delta T_{\rm diss, 2} = 0 \,,\\ 
& \Delta T_{\rm ad, 1} + \Delta T_{\rm ad, 2} = 0 \,. 
}
In another word, there is no change in $T_{\rm diss}$ or $T_{\rm ad}$ after averaging over gas elements.

In comparison, the entropy variation for each element
\eqs{
\Delta S & = \int_{\rho_{\rm i}} ^{\rho_{\rm f}} \dd \ln K(\rho) \\
& = - \gamma \ln \br{\frac{\rho_{\rm f}}{\rho_{\rm i}}} \\
& = \gamma \bb{\ln(T_{\rm f}) - \ln(T_{\rm i})} \,.
}
It is proportional to the change in $\ln T$ rather than that in $T$.
Thus, $\Delta S_1 + \Delta S_2 = \gamma \bb{2\ln(T_{\rm f}) -  \ln(T_{\rm i, 1}) - \ln(T_{\rm i, 2})} > 0$, i.e., entropy of the combined system increases during mixing.

\section{Control of mixing in the simulation}
\label{app:mixing_control}
\begin{figure}
\centering
    \includegraphics[width=.37\textwidth]{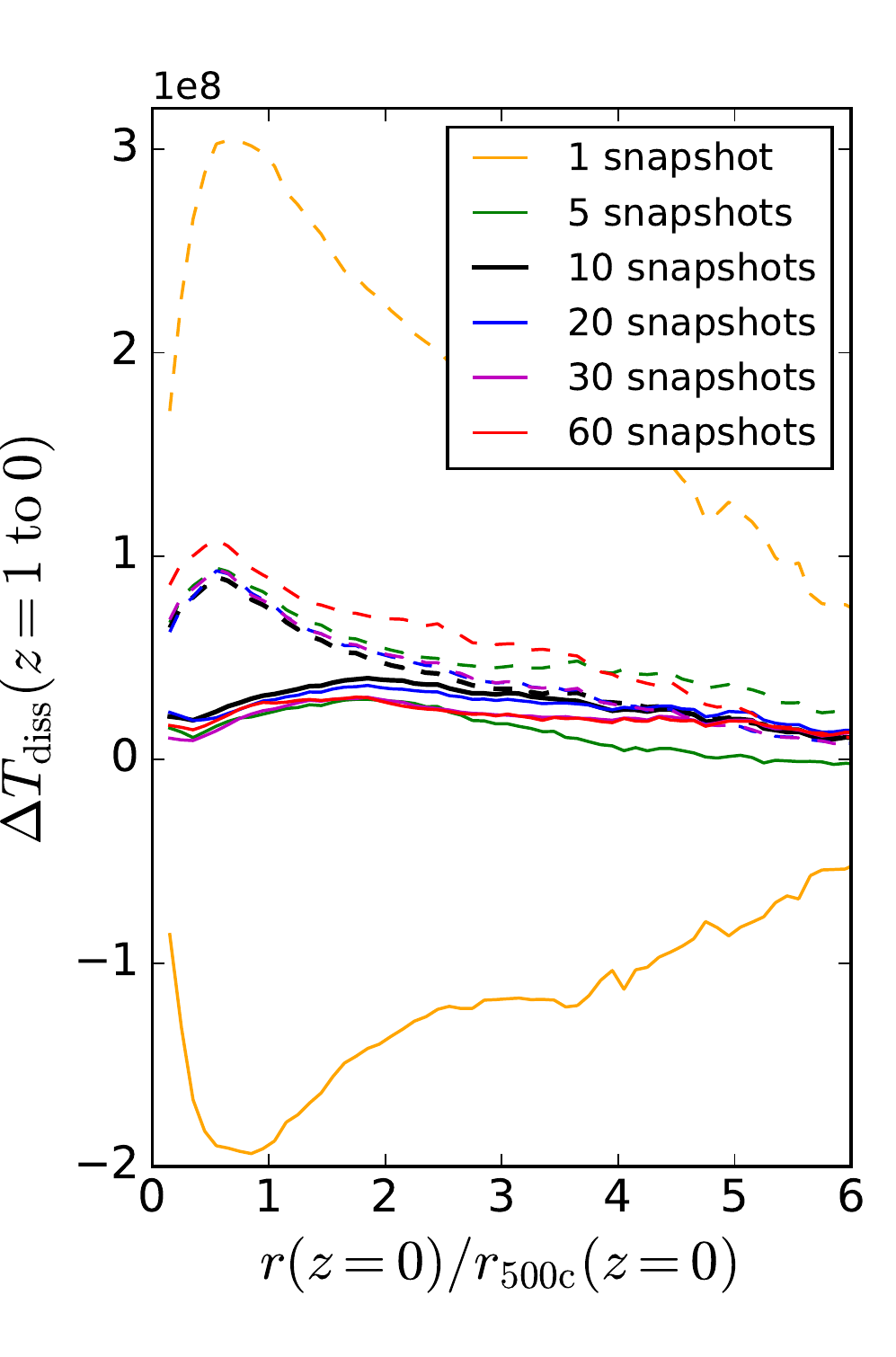}
  \caption{Same as Fig.\;\ref{fig:heating_region} but showing the effect of different binning sizes in time when distinguishing compression/rarefaction regions. The solid lines represent the contribution from compression regions, while the dashed lines represent that from rarefaction regions.}
\label{fig:heating_region_binning}
\end{figure}
Since no viscosity is implemented in the simulation, mixing happens via numerical viscosity and its amount is resolution-dependent. Thus,
the mixing in a Eulerian simulation partly reflects a physical effect, and partly a numerical effect. There have been studies discussing the extent to which this mixing is physical in the context of entropy cores in galaxy clusters \citep[e.g.,][]{vazza11b}. When re-simulating the original Eulerian simulation using Lagrangian tracer particles, additional spurious mixing of particles is introduced due to the finite spatial and time resolution of the velocity field used to advect the tracer particles. These mixing processes affect our analysis in three ways. We explain these effects as well as how we control them in the following.

Regardless of its origin, mixing leads to exchanges of $T_{\rm diss}$ and $T_{\rm ad}$ among neighboring gas elements. This is a source of noise for our analysis whose purpose is to study ICM heating during the growth of a galaxy cluster. However, as shown in Appendix.\;\ref{app:mixing_phys}, the mixing effect in $T_{\rm diss}$ and $T_{\rm ad}$ naturally disappears after sufficient spatial averaging. This is why the $r-t$ distribution of the dissipative heating rate (Fig.\;\ref{fig:dTdiss_diffuse}) is not much affected, that we do not see many $\dd r- \dd t$ bins with negative $\dd T_{\rm diss}$.

The analysis that is most affected by the mixing effect is the decomposition of the dissipative heating $\Delta T_{\rm diss}$ into that occurred in compression / rarefaction regions (Fig.\;\ref{fig:heating_region} and Section\;\ref{sec:heating_region}). Isobaric mixing introduces an anti-correlation between $\dd K$ and $\dd \rho_{\rm gas}$, and thus boosts the amount of $\dd T_{\rm diss}$
that occurs in rarefaction regions and reduces that in compression regions. Again, binning is needed to suppress this contamination by suppressing the mixing effect. In Section\;\;\ref{sec:heating_region}, we bin the tracer particle density variations in time when determining whether the particle resides in a compression region or a rarefaction region; i.e., we take the density difference between snapshots with a larger time difference rather than that between the subsequent snapshots. The bin size should be large enough to sufficiently suppress the contamination, but small enough to avoid over-smoothing the two regions. We show how the lines in Fig.\;\ref{fig:heating_region} is influenced by the bin size in Fig.\;\ref{fig:heating_region_binning}. Without binning (orange lines), the contamination dominates the outcome, causing the total $\Delta T_{\rm diss}$ in the compression region to be negative. Binning a few snapshots already largely remove this contamination. For a binning of 10-30 snapshots, the results are rather stable to the size of the binning, showing consistent results of an increasing ratio of $\Delta T_{\rm diss}$ in compression regions over that in rarefaction regions. With a bin size of 60 snapshots (red lines), the over-smoothing effect becomes significant, which reduces $\Delta T_{\rm diss}$ in the compression region due to smeared shock fronts. This effect already shows up slightly when the bin size increases from 10 to 30 snapshots. Therefore, in Fig.\;\ref{fig:heating_region} we have chosen the bin size to be 10 snapshots where the overall contamination from mixing and over-smoothing is minimized.

When a mixing process is temporally not well-resolved in the simulation, it can lead to large jumps in the thermodynamical quantities of a gas element between subsequent time steps in locations with large density gradients. In this case, if we still compute $\dd T_{\rm diss}$ and $\dd T_{\rm ad}$ according to Eqs.\;\ref{eq:dTdiss} and \ref{eq:dTad} , they are
\eq{
\dd T_{\rm diss} = T_{\rm i} \dd \ln K = T_i \dd S = \gamma T_i \bb{\ln(T_{\rm f}) -  \ln(T_{\rm i})}\,,
}
and
\eq{
\dd T_{\rm ad} = (\gamma - 1) T_{\rm i} \dd \ln \rho = - (\gamma - 1) T_i \bb{\ln(T_{\rm f}) -  \ln(T_{\rm i})}\,,
}
i.e., they are proportional to the change in $\ln T$. Therefore, like the entropy of the combined system in the example presented in Appendix.\;\ref{app:mixing_phys}, there are net changes in $\dd T_{\rm diss}$ and $\dd T_{\rm ad}$ even after averaging, positive for $\dd T_{\rm diss}$ and negative for $\dd T_{\rm ad}$. As this contamination typically occurs at the boundary between diffuse ICM and dense structures such as filaments and clumps, we limit it by restricting all analysis in this paper to diffuse regions. This measure also reduces the other mixing effect described above.

\end{document}